\documentclass[aps,prb,twocolumn,reprint,superscriptaddress,showkeys,floatfix,amsmath,amssymb]{revtex4-2}
\usepackage{epsfig}
\usepackage{color}
\usepackage{booktabs}
\usepackage[colorlinks,allcolors=blue,pdftex]{hyperref}
\usepackage{graphicx}
\usepackage{hypernat}
\usepackage{csquotes}
\usepackage{xcolor, soul}
\usepackage{dcolumn}
\usepackage{bm}
\usepackage{natbib}

\begin{document}

\preprint{APS/123-QED}

\title{ Newly discovered magnetic phase: A brief review on Altermagnets}

	\author{R. Tamang}
	\affiliation{Advanced Computation of Functional Materials Research Lab (ACFMRL) Department of Physics, Mizoram University, Aizawl-796004, India}%
	\affiliation{Physical Sciences Research Center (PSRC), Department of Physics, Pachhunga University College,  Aizawl-796001, India}

	\author{Shivraj Gurung}
		\affiliation{Physical Sciences Research Center (PSRC), Department of Physics, Pachhunga University College,  Aizawl-796001, India}

	\author{D. P. Rai}
	\email[D. P. Rai:]{dibyaprakashrai@gmail.com}
	\affiliation{Advanced Computation of Functional Materials Research Lab (ACFMRL) Department of Physics, Mizoram University, Aizawl-796004, India}%
	\affiliation{Peter Gr\"unberg Institute, Forschungszentrum J\"ulich and JARA, J\"ulich, Germany}
	\homepage{www.mzu.edu.in}
    
 \author{Samy Brahimi}
  \affiliation{Peter Gr\"unberg Institute, Forschungszentrum J\"ulich and JARA, J\"ulich, Germany}
 \affiliation{Laboratoire de Physique et Chimie Quantique, Universite Mouloud Mammeri de Tizi-Ouzou, 15000 Tizi-Ouzou, Algeria}

 \author{Samir Lounis}
 \email[Samir Lounis:]{s.lounis@fz-juelich.de}
 \affiliation{Peter Gr\"unberg Institute, Forschungszentrum J\"ulich and JARA, J\"ulich, Germany}
 \affiliation{Faculty of Physics, University of Duisburg-Essen and CENIDE, Duisburg, Germany}





\date{\today}

\begin{abstract}
Recently, a new magnetic phase, termed altermagnetism, has caught the attention of the magnetism and spintronics community. This newly discovered magnetic phenomenon differs from traditional ferromagnetism and antiferromagnetic. It generally lacks net magnetization and is characterized by unusual non-relativistic spin-splitting and broken time-reversal symmetry. This leads to novel transport properties such as the anomalous Hall effect, the crystal Nernst effect, and spin-dependent phenomena that cannot be fully explained by traditional magnetic theories. Spin-dependent phenomena such as spin currents, spin-splitter torques, and high-frequency dynamics emerge as key characteristics in altermagnets. This paper reviews the main aspects pertaining to altermagnets by providing an overview of theoretical investigations and experimental realizations.
We discuss the most recent developments in altermagnetism, its comparison to other magnetic orders, and future prospects for exploiting its unique properties in next-generation devices.
\end{abstract}

\maketitle

\section{Introduction}
\begin{figure*}[hbtp]
	\centering
\includegraphics[width=0.30\linewidth]{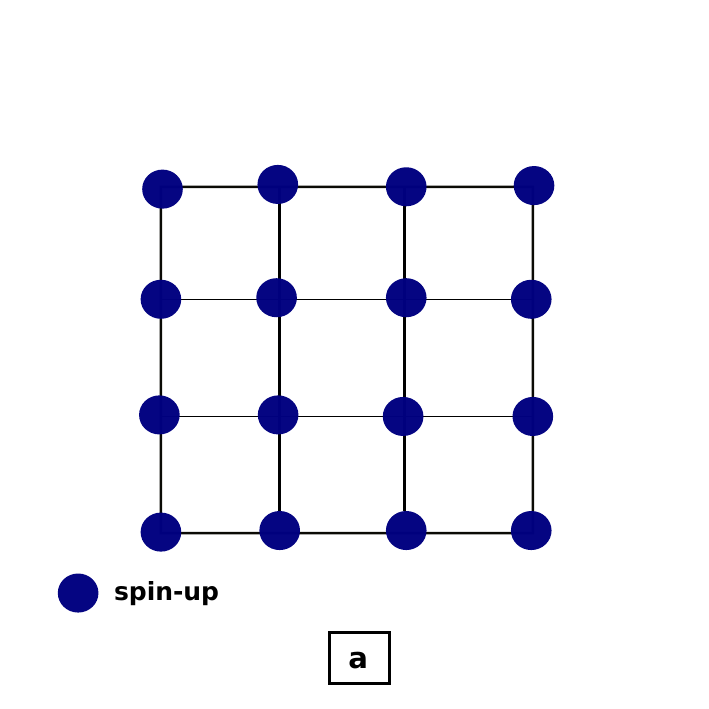}
\includegraphics[width=0.30\linewidth]{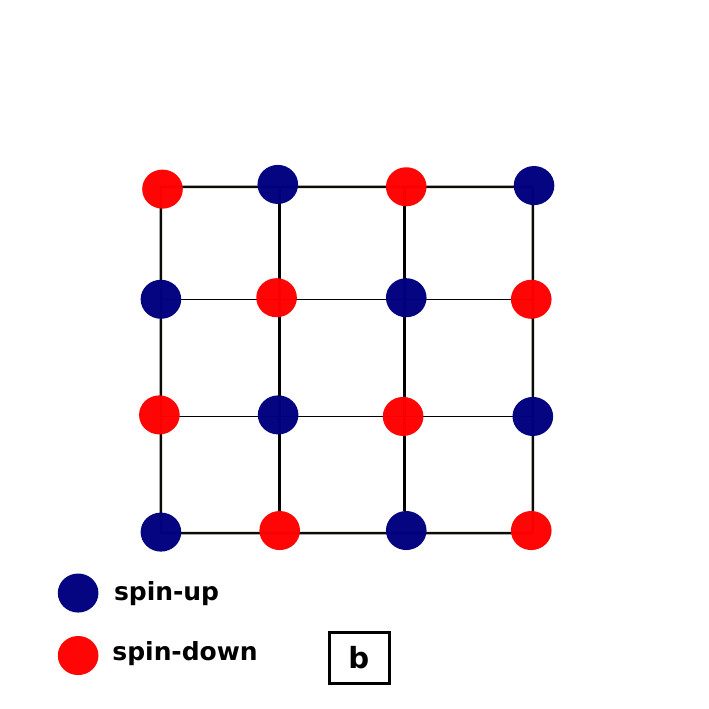}
\includegraphics[width=0.30\linewidth]{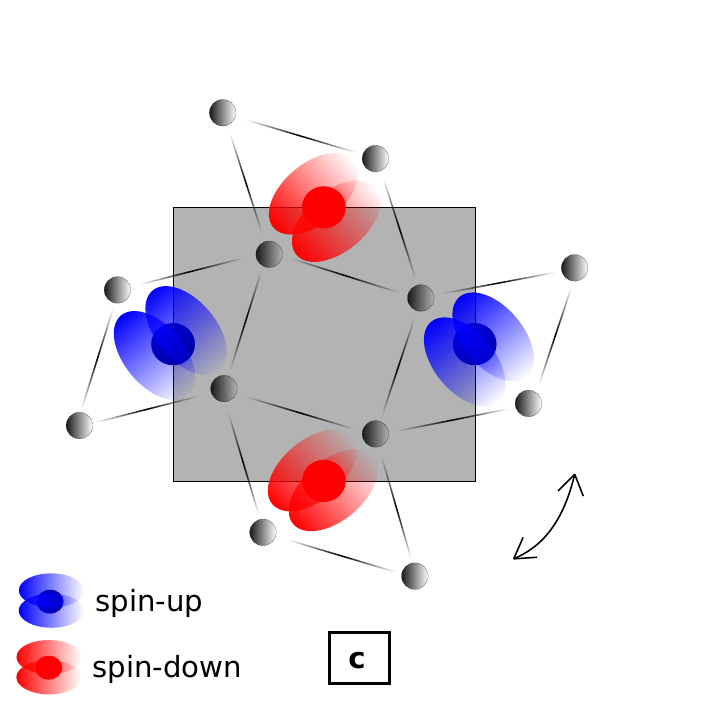}
\caption{a) Ferromagnetic state, compared to an b) antiferromagnetic configuration with compensated magnetic ordering. c) Altermagnetic RuO$_2$ with d-wave symmetry of the magnetic Ru (red and blue) atoms and the non-magnetic O (grey) atoms, the double-headed arrow indicates that two opposite spin sub-lattices are connected by rotation only.}
	\label{fig:fig1}
\end{figure*}

\begin{figure}[hbtp]
	\centering
	\includegraphics[width=1.0\linewidth]{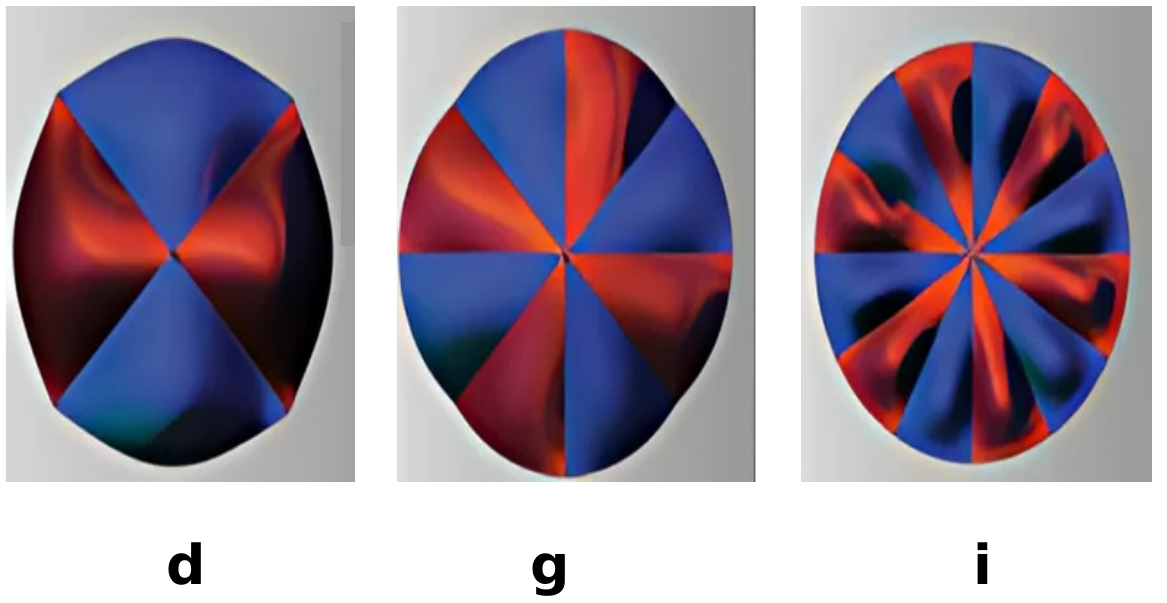} 
	\caption{Schematic d-,g- and i- even-parity waveform of altermagnets. The figure is adapted from ref\cite{vsmejkal2022beyond} under  CC-BY 4.0 International license. Published by American Physical Society(APS)}
	\label{fig:waveform}
\end{figure}

\begin{figure*}[hbtp]
	\centering
    \includegraphics[width=1.0\linewidth]{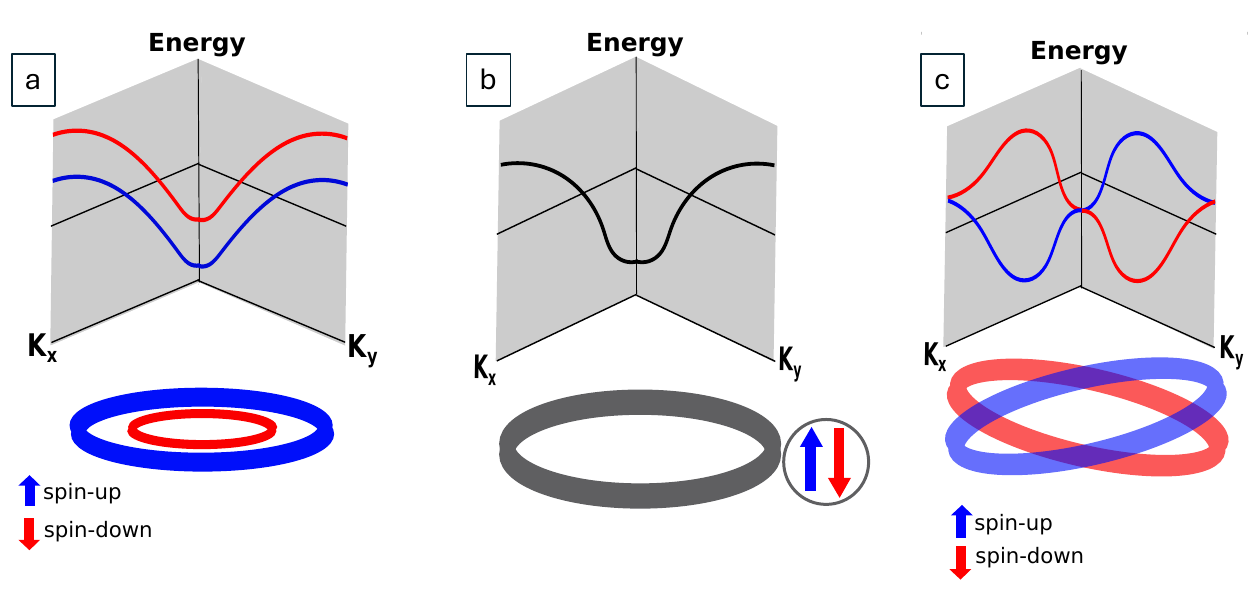}
	\caption{a) Band diagram of ferromagnetic materials corresponding to spin-up (blue) and spin-down (red) configuration followed by constant splitting of energy iso-surface depicted in the bottom of the diagram b) Band diagram corresponding antiferromagnetic materials with degenerate band structure of spin-up and spin-down configuration. c) Band diagram corresponding altermagnetic materials showing significant alternating spin-splitting. These figures are reproduced from Ref.\cite{vsmejkal2022beyond,vsmejkal2022emerging} under CC BY 4.0 International license. Published by APS, copyright 2022.}
	\label{fig:band_split}
\end{figure*}

\begin{figure*}[h]
	\centering
\includegraphics[scale=1.4]{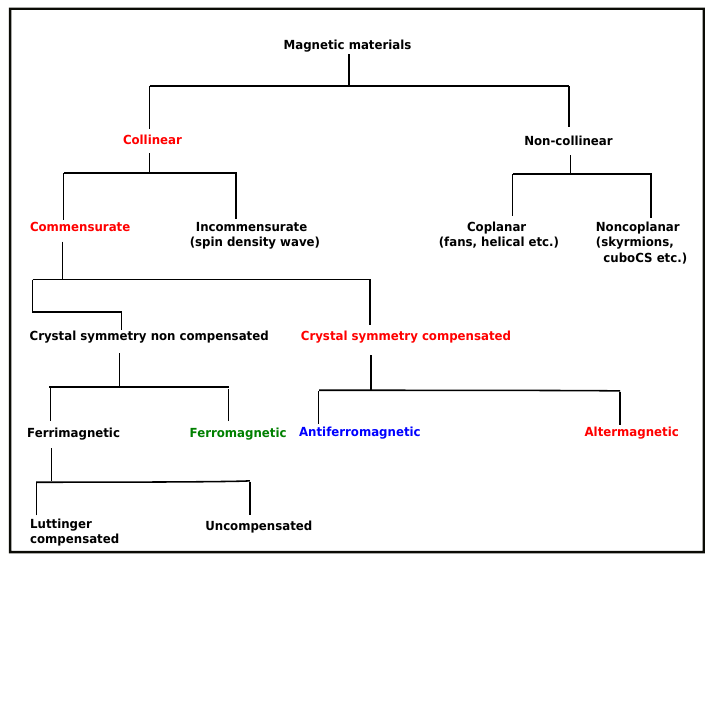} 
	\caption{Altermagnetism depicted as a new type of magnetic order in a categorization scheme comprising multiple magnetic orders. The figure is reproduced from ref\cite{mazin2022altermagnetism} under  CC-BY 4.0 International license. Published by American Physical Society(APS), copyright 2022.}
	\label{fig:table}
\end{figure*}

A new magnetic phase known as altermagnetism (AM) has been recently coined\cite{vsmejkal2022beyond} apart from common ferromagnetism (FM) and antiferromagnetism (AFM). The two main phases of magnetism have previously been understood as follows: antiferromagnetism, in which spin alignment is antiparallel and cancels out net magnetization, and ferromagnetism characterized by parallel spin alignment producing net magnetization \cite{neel1953some}[Fig.\ref{fig:fig1}(a,b)]. Altermagnetism, however, is a new phenomenon that combines the aspects of both the conventional phases, i.e., FM and AFM, without precisely falling into either category. Altermagnets are designated by broken time-reversal symmetry (TRS) and display unconventional non-relativistic spin-splitting (NRSS) analogous to even-parity d-, g-, or i-wave symmetry [Shown in Fig.\ref{fig:waveform} \cite{vsmejkal2022beyond,vsmejkal2022emerging}]  without requiring relativistic spin-orbit coupling (SOC).
 Altermagnets have unique characteristics that combine the ultrafast dynamics and external magnetic field endurance of antiferromagnets with the substantial transport and optical effects often associated with ferromagnets\cite{shao2021spin}. 
The arrangement of the local environment formed by the bonding between magnetic and non-magnetic atoms in altermagnets plays a pivotal role in shaping their spin-splitting properties (as shown in Fig.\ref{fig:fig1}(c))\cite{gomonay2024structure}. The d-wave altermagnets can be regarded as magnetic counterparts to unconventional d-wave superconductors\cite{vsmejkal2022beyond} and as realizations of the nematic state in both real and spin spaces. 
Based solely on the symmetry of the spin densities in the crystal lattice, altermagnetism can be realized in various metallic and insulating materials\cite{vsmejkal2022emerging}. It results from the electron-electron interactions and single-particle potentials of the crystal lattice\cite{jungwirth2024altermagnets}.
This phenomenon is different from the spin-polarized even-parity wave Pomeranchuk instability\cite{wu2007fermi,pomeranchuk1958stability}, which is a purely electronic instability of the correlated Fermi liquid. However, the resulting symmetry of the magnetic ordering in the momentum space is similar. The model bands are used to compare three magnetic phases, viz. FM, AFM and AM are represented in Fig.\ref{fig:band_split}. In the classification of magnetic materials, Mazin et al. classified altermagnets as: \enquote{collinear} and \enquote{commensurate} with \enquote{symmetry-compensated} \cite{mazin2022altermagnetism} (as shown in Fig.\ref{fig:table} ). Hence, they share similar origins with their counterpart antiferromagnets; however, they differ significantly in certain aspects, such as NRSS and broken TRS.


\begin{figure*}[h]
	\centering
	\includegraphics[scale=1.40]{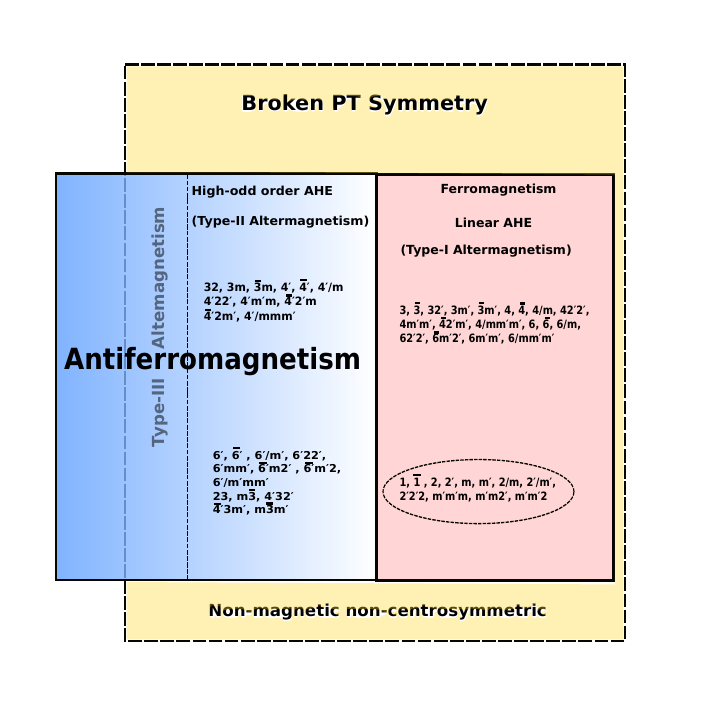} 
	\caption{Classification of multiple magnetic states based on magnetic point group: Any magnetic states that do not fall into the ferromagnetic point group are antiferromagnetic (faded blue box), and all ferromagnets (pink box) are members of this ferromagnetic point group. All ferromagnets show linear AHE, some antiferromagnets show high-odd-order AHE, and all magnetic states with odd-order AHE have broken $\mathcal{PT}$ symmetry.
The dashed ellipse contains magnetic point groups applicable to type-I altermagnetism proposed for collinear spins; type-I altermagnets are all members of the ferromagnetic point group. Type-III altermagnets cannot display any odd-order AHE along the principal axes, but all type-II altermagnets are capable of displaying high-odd-order AHE. Non-magnetic and non-centrosymmetric states are indicated by the yellow area. This figure is reproduced from Ref.\cite{cheong2024altermagnetism} under CC-BY 4.0 International license. Published by Springer Nature, copyright 2024.}
	\label{fig:fig2}
\end{figure*}

\begin{figure*}[h]
	\centering
\includegraphics[scale=1.4]{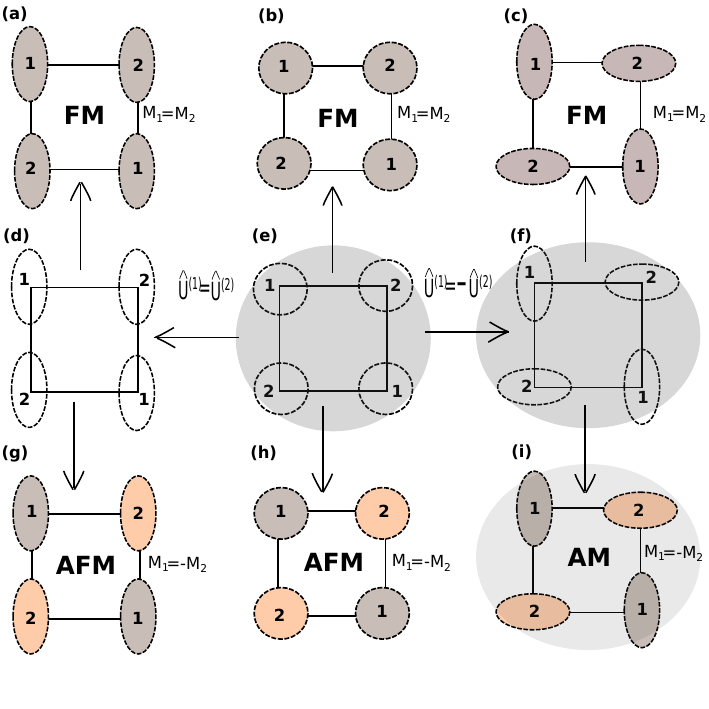} 
	\caption{(a to c) represent the ferromagnetic phases (FM with M$_1$=M$_2$) while  
(g to i) represent the staggered magnetic phases (AFM or AM with M$_1$=-M$_2$) and (d,f) are structural phases. 
Any magnetic phase can be derived from the highly symmetric hypothetical phase demarcated in (e) with no magnetic order and isotropic local environment of magnetic atoms (shown with circles). The distortion of local environment leads to two different structural phases with equal ($\hat{U}^{(1)}$=$\hat{U}^{(2)}$) or distorted($\hat{U}^{(1)}$=$-\hat{U}^{(2)}$) local environment ($U$'s are second rank symmetric tensor illustrating feasible distortion of the sublattices). Combining the staggered environment with the staggered order gives altermagnetism. The sign of the exchange coupling between atoms 1 and 2 determines whether the magnetic order results in staggered (M$_1$$=$-M$_2$ ) or ferromagnetic (M$_1$ = M$_2$)  structures. This figure is reproduced from Ref.\cite{gomonay2024structure} under CC-BY 4.0 International License. Published by Springer Nature, copyright 2024.  }
	\label{fig:example1}
\end{figure*}

According to Cheong et al.\cite{cheong2024altermagnetism}, altermagnets can be divided into three categories: type I, II, and III. Previously, AM was assumed to be limited to only\enquote{collinear} antiferromagnets with distorted local environments. However, they proposed that, in general, altermagnets can encompass both collinear and non-collinear antiferromagnets; one predicted example of the non-collinear order is YMn0$_3$ \cite{brown2006neutron}. The altermagnets of type I possess a net magnetic moment and display ferromagnetic characteristics even when external influences are absent. In contrast, II- and III-type altermagnets have no net magnetic moment. However, Type II altermagnets can demonstrate a non-zero net magnetic moment and ferromagnetic behavior when subjected to external perturbations that are $\mathcal{PT}$-symmetric ($\mathcal{PT}$ represents the interplay of parity and time reversal symmetry), including influences like an applied electric field, thermal current, illumination from light, or mechanical stress. Type-I altermagnets are members of the ferromagnetic point group
shown in Fig.\ref{fig:fig2} by dashed ellipse and exhibit linear anomalous Hall effect(AHE), while type-II altermagnets belong to an antiferromagnetic point group and possess high-odd-order AHE. However type-III altermagnets do not exhibit a strong odd-order AHE.\cite{cheong2024altermagnetism} In the context of realizing altermagnets, the systematic distortion of the local environment from a hypothetical high-symmetry phase can lead to distorted phases depending on the second rank symmetric tensor $\hat{U}$'s, potentially resulting in altermagnetism when the magnetization is compensated as illustrated in Fig.\ref{fig:example1}.

\begin{figure}[hbtp]
	\centering
	\includegraphics[width=0.7\linewidth]{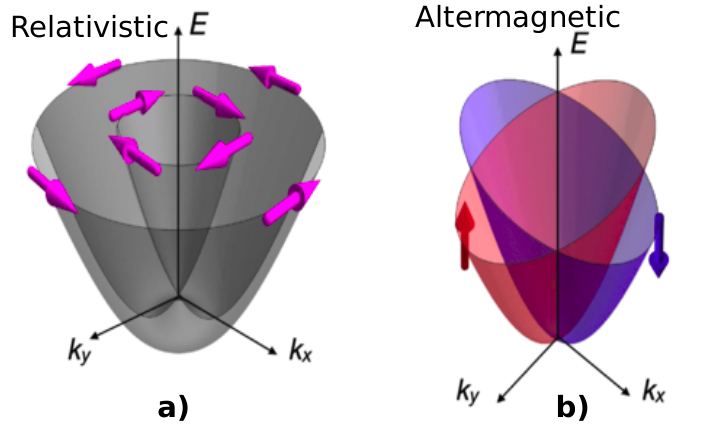} 
	\caption{Schematic a) SOC induced relativistic spin-splitting of bands within the Rashba model. b) non-relativistic d-wave spin-splitting of bands. The figures are adapted from Ref.\cite{vsmejkal2022emerging}under CC-BY 4.0 International license. Published by APS, copyright 2022.}
	\label{fig:rashba}
\end{figure}

 Besides the aforementioned Anomalous Hall effect emerging in altermagnets, which promotes the latter to be used for potential spintronics applications, other intriguing phenomena were found. For example, spin-currents \cite{shao2021spin,gonzalez2021efficient,bose2022tilted}, spin-splitter torques\cite{bai2022observation,karube2022observation}, and giant magnetoresistance\cite{vsmejkal2021giant} effects. Altermgnets can support high-frequency dynamics(THz)\cite{baltz2024emerging,qiu2023terahertz}, robust spin currents, and distinct topological states\cite{reichlova2024role,zhu2023topological,li2024creation}. In general, these properties promote altermagnets to be promising materials for information technology\cite{baltz2024emerging}. Spintronic devices would switch faster and operate at lower power compared to conventional charge-based electronics\cite{kang2015spintronics,joshi2016spintronics}. 
 Theoretically, reported g-wave altermagnets have chemical stoichiometry ABX$_3$ e.g. CsCoCl$_3$, RbCoBr$_3$, and BaMnO$_3$. In the aforementioned materials, the spin-splitting is much larger in comparison to the one induced by SOC\cite{mcclarty2024landau}. The distinctive difference between NRSS and Rashba spin splitting is shown in Fig.\ref{fig:rashba}. Advanced experimental methods such as neutron scattering, magnetic resonance, and advanced spectroscopic techniques can probe the unique spin structures of altermagnets. First-principle computations demonstrated NRSS in various materials, with about more than 200 compounds predicted as candidate altermagnets\cite{bai2024altermagnetism}. Using a machine learning-based approach, potential candidates have been identified based on their electronic structures\cite{gao2023ai}. However, only a limited number have been experimentally investigated (as shown in Table \ref{tab:example}), resulting in the discovery of only a few material candidates. This paves the way for exploring a wide range of 2D and bulk materials that could potentially host this novel magnetic phase. 

  Realizing altermagnetism in 2D can involve twisted antiferromagnetic bilayers\cite{he2023nonrelativistic}, particularly through Van der Waals stacking. Pan et al.\cite{pan2024general} proposed a General Stacking Theory (GST) to predict altermagnetism in bilayer based on the symmetry of the monolayer components, subject to the constraint that there must be an antiferromagnetic coupling between the two monolayers. A notable extension of this framework hinges on the exploration of direct stacking of monolayers without twisting. A stacking operator ($\hat{P}$) transforms a monolayer (L) into another layer (L$^{\prime}$) by applying a series of symmetry operations, translations, or rotations. Together, L and L{$^{\prime}$} constitute the bilayer (B) generated directly\cite{pan2024general}.
 
  It is seen that strain may influence crystal symmetry in some materials, leading to a phase transition from AFM to AM. The bulk ReO$_2$ prefers the $\alpha$ monoclinic phase showing antiferromagnetic behaviour, which, under the application of pressure changes to tetragonal R-ReO$_2$ that is altermagnetic. Murnaghan fitting data show that R-ReO$_2$ is comparatively stable compared to monoclinic ReO$_2$\cite{chakraborty2024strain}. Traditional studies, such as those of Rashba and Dresselhaus\cite{rashba1960spin,dresselhaus1955spin}, demonstrated the effects of SOC in non-centrosymmetric systems that are restricted to high-Z compounds, which are chemically less stable\cite{yuan2020giant}. There is a significant need for alternative mechanisms, particularly in antiferromagnetic (AFM) materials, to enable spin splitting in thermodynamically stable, low-Z compounds. Past research on SOC-induced spin splitting has focused primarily on high-Z materials, leaving low-Z AFM systems largely underexplored.\cite{yuan2020giant}.

Based on first-principles, Sattigeri et al.\cite{sattigeri2023altermagnetic} reported the emergence of surface states exhibiting altermagnetic behaviour with
spin-splitting is believed to be highly dependent on magnetic space groups. The properties of the magnetic surface state have been investigated in three exemplary space groups: orthorhombic ($Pbnm$(62)), hexagonal ($P6_3/mmc$(194)), and tetragonal ($P4_2/mnm$ (135)). It was found that the altermagnetic properties on certain surfaces are preserved but annihilated on others because of certain spin-splitting characteristics in the Brillouin zone (BZ). For example in, LaMnO$_3$(Space group-62) belonging to the perovskite family with A-type antiferromagnetism, surfaces (010) and (001) are devoid of AM whereas the (100) surface hosts AM with spin splitting of about 30 meV. Furthermore, it was demonstrated that the electric field can have significant control over the surface states and triggers altermagnetism on the otherwise blind surface\cite{sattigeri2023altermagnetic}.\\

\section{Discussion}
In traditional antiferromagnets, the band structure is characterized by Kramer's degeneracy since sublattice exchanging symmetry is translational or parity combined with time reversal; hence, the symmetry is preserved. Alternatively, in the case of altermagnets, the sublattices aligned in an anti-parallel manner are linked solely through a rotation, whether that be proper or improper, as well as symmorphic or non-symmorphic\cite{vsmejkal2022emerging}. 

As mentioned above, the ab initio calculation performed on the rutile RuO$_2$ revealed an unanticipated d-wave spin splitting of the Fermi surface\cite{vsmejkal2020crystal}. A recent study suggests that the coexistence of staggered antiferromagnet and orbital order(OO) is capable of producing robust altermagnetism, giving rise to significant spin-splitting and spin conductivity. Electron correlations can generate a phase in which Neel AFM and staggered OO coexist, producing a d-wave altermagnetic phase\cite{leeb2024spontaneous}. In monolayer, RuF$_4$, the magnetic properties of ruthenium are shaped by the effect of the crystal field of negative fluorine ions and the strong hybridization between the orbitals Ru 4d and F 2p. The crystal field splits the Ru 4d orbitals into t$_{2g}$ and e${_g}$ orbitals, typically suggesting a specific magnetic moment of 2$\mu_B$. However, density functional theory (DFT) calculations by Milivojevic et al. revealed a magnetic moment of 1.464$\mu_B$, lower than expected, due to the predicted significant hybridization of Ru 4d and F 2p orbitals. Moreover, the inclusion of SOC in the calculations introduces weak ferromagnetism, arising from the slight canting of the Ru spins\cite{milivojevic2024interplay}. 
There are specific types of antiferromagnetic materials that exhibit non-relativistic spin splitting, known as the fourth spin splitting prototype SST-4\cite{yuan2020giant,yuan2021prediction}. SST-4 encompasses both non-collinear and collinear antiferromagnets, the collinear antiferromagnets that are classified under SST-4 feature motif pairs, each pair carries opposite magnetic moments that are not linked by translation or inversion. In contrast, altermagnets also require spin-structure motif pairs that are not connected by translation or inversion, but can be linked through proper or improper spatial rotation operations. Yuan et al.\cite{yuan2024non} proposed that there exist NRSS antiferromagnets containing spin-structure motif pairs that are not connected by any type of rotational symmetry, whether proper or improper, representing a new kind of NRSS antiferromagnets that are not classified as altermagnets. This specific group of antiferromagnets exhibits splitting even at the $\Gamma$ point, a phenomenon absent in altermagnets. Examples of such compounds include ternary magnetic nitrides such as MnXN$_2$ (where X=Si, Ge, or Sn) and their cation-order variations. These compounds have the potential to produce spin currents without cancellation because of alternating spin polarization. 

\subsection{Anomalous hall effect}
 
The AHE arises from the breaking of time-reversal symmetry in magnetic materials, typically due to their intrinsic magnetization and spin-orbit coupling\cite{nagaosa2010anomalous}. Therefore, the AHE is also coined as the spontaneous Hall effect, which surprisingly was predicted and observed to occur in non-collinear antiferromagnets\cite{chen2014anomalous,kubler2014non,nakatsuji2015large,surgers2016anomalous,boldrin2019anomalous,vsmejkal2022anomalous} and, particularly in altermagnets\cite{attias2024intrinsic,leiviska2024anisotropy,feng2022anomalous,tschirner2023saturation,jin2024anomalous}.  
The spontaneous Hall voltage is a phenomenon in which specific internal magnetic configurations cause the electrons to gain transverse velocity. This effect is related to the antisymmetric dissipation-free portion of the conductivity tensor, which is represented by the Hall pseudovector and controls the Hall current\cite{vsmejkal2020crystal}. 
In collinear altermagnets, the simplified magnetic structure defined solely by the spin orientations and spatial arrangement of magnetic atoms does not produce a spontaneous Hall conductivity. The necessary asymmetry emerges only when additional atoms, often nonmagnetic, occupy noncentrosymmetric sites. For example, in RuO$_2$, the arrangement of oxygen atoms induces an asymmetry in the magnetization density between the two opposing Ru spin sublattices. This breaks the combined symmetry of time-reversal and lattice translation associated with altermagnets.~\cite{vsmejkal2020crystal}, which leads to an additional Hall effect, named the crystal Hall effect, that is independent of the net magnetization or presence of SOC, which highlights a key advantage of altermagnets \cite{betancourt2023spontaneous,reichlova2024macroscopic,mcclarty2024landau}. After initial predictions of a significant AHE in RuO$_2$\cite{vsmejkal2020crystal},  experiments followed quickly with a confirmation \cite{feng2022anomalous,tschirner2023saturation}. 
 
Furthermore, AHE has been observed in altermagnetic Mn$_5$Si$_3$\cite{leiviska2024anisotropy}  and hexagonal MnTe\cite{kluczyk2024coexistence}. In the thin film of RuO$_2$, one can observe the reorientation of the N\'eel vector from [001] easy plane to [110] hard plane caused by the application of electric field, allowing for the observation of AHE
\cite{vsmejkal2022anomalous,tschirner2023saturation,sattigeri2023altermagnetic}.

The detection of the Hall effect in antiferromagnets provides new possibilities for designing spintronic devices that take advantage of these materials' robustness and unique features. Potential applications include high-speed, low-power, and low-dissipative electrical devices as well as innovative quantum computer elements\cite{khalili2024prospects,jungwirth2016antiferromagnetic,vsmejkal2018topological,baltz2018antiferromagnetic}.

\subsection{Nernst effects}
The Nernst effect is a fundamental phenomenon that emerges in a longitudinal temperature gradient, resulting in a transverse voltage without the application of an external magnetic field. Similar to the AHE, in conventional collinear antiferromagnets the anomalous Nernst effect (ANE) and the anomalous thermal Hall effect (ATHE) are expected to vanish. 
In work conducted by Zhou et al.\cite{zhou2024crystal}, however, it was found that significant thermal transport effects exist in RuO$_2$, specifically known as the crystal Nernst effect (CNE) and the crystal thermal Hall effect (CTHE). These effects resemble the ANE and ATHE but are unique to the crystal structure of RuO$_2$, where the local environment of the nonmagnetic atom is anisotropic (similar to the discussion of the previous subsection on the AHE) and leads to the emergence of altermagnetism\cite{vsmejkal2022emerging}. The study suggests that the sporadic thermal and electrical transport coefficients in RuO$_2$ adhere to an extended Wiedemann-Franz law in a wide temperature range of (0-150K)\cite{zhou2024crystal}, a range much wider than what is typically expected for traditional magnetic materials.

As a result, altermagnetic RuO$_2$ could play a notable role in the realm of spin caloritronics due to its distinct thermal transport properties not attainable by conventional ferromagnets and antiferromagnets. The CNE in RuO$_2$ shows high anisotropy, which means that its intensity and direction are heavily influenced by the alignment of the N\'eel vector and can be experimentally observed in films oriented along the directions [110] or [001]. 

Within RuO$_2$, Weyl fermions\cite{hasan2021weyl} contribute substantially to Berry curvature. The latter behaves like a magnetic field in the k-space, which impacts electron motion and subsequently affects various transport properties. Weyl fermions are quasi-particles that materialize in substances where electronic bands intersect at discrete points known as Weyl points\cite{bradlyn2016beyond}. The existence of Weyl fermions leads to significant Berry curvature effects, which are crucial in explaining irregular thermal transport events such as CNE and CTHE. Other sources of Berry curvature include pseudonodal surfaces and ladder transitions. Anomalous transport coefficients can be obtained by integrating the Berry curvature with the BZ\cite{zhou2024crystal,vsmejkal2020crystal}.


\subsection{Experimental techniques and theoretical approaches to explore altermagnetism}
 
 Various sophisticated methods have been utilized to demonstrate the existence of altermagnets.  
\begin{itemize}
    \item Neutron scattering techniques: used for RuO$_2$, $\alpha$-MnTe\cite{kessler2024absence,lovesey2023templates}.

    \item XMCD: X-ray magnetic circular dichroism applied on MnTe, MnF$_2$ \cite{hariki2024x,hariki2024determination}.

    \item Muon spin rotation/relaxation (${\mu}$SR): utilized on RuO$_2$ \cite{kessler2024absence}. 

	\item ARPES and Spin SX-ARPES: Angular-resolved photoemission spectroscopy (APRES) and spin-integrated soft X-ray (SX) ARPES experiments were utilized to directly detect spin splitting in various altermagnets\cite{lee2024broken,liu2024absence,krempasky2024altermagnetic,osumi2024observation,zhu2024observation,reimers2024direct,ding2024large,lin2024observation}. 
	
	\item DFT (Density Functional Theory): First-principles calculations were performed to provide theoretical predictions to motivate and compare with experimental findings, focusing on various materials including RuO$_2$\cite{vsmejkal2020crystal,lin2024observation,ahn2019antiferromagnetism}, CrSb\cite{reimers2024direct,ding2024large}, MnTe\cite{lee2024broken,krempasky2024altermagnetic,osumi2024observation}, MnF$_2${\cite{bhowal2022magnetic,yuan2021prediction}}, and others \end{itemize}.
     
     \begin{table}[h!]
    \caption{NRSS on various altermagnet materials from the experiment.}
    \begin{ruledtabular}
    \begin{tabular}{ccccc}
        Materials & Space Group& Experiment & Reports & Reference \\ \hline
        CrSb      & $P6_3/mmc$      & ARPES      & NRSS detected     & \cite{reimers2024direct,ding2024large}      \\
        MnTe      & "     & "      & "     &  \cite{lee2024broken,krempasky2024altermagnetic,osumi2024observation}     \\
        MnTe$_2$  & $Pa\bar{3}$      & "      & "      & \cite{zhu2024observation}      \\
        RuO$_2$   & $P4_2/mnm$      & "     & "     & \cite{lin2024observation}     \\
        RuO$_2$   & $P4_2/mnm$      & "     & NRSS undetected     & \cite{liu2024absence}     \\
        
    \end{tabular}
    \end{ruledtabular}
    \label{tab:example}
\end{table}
     
\begin{figure}[h]
	\centering
	\includegraphics[width=\linewidth]{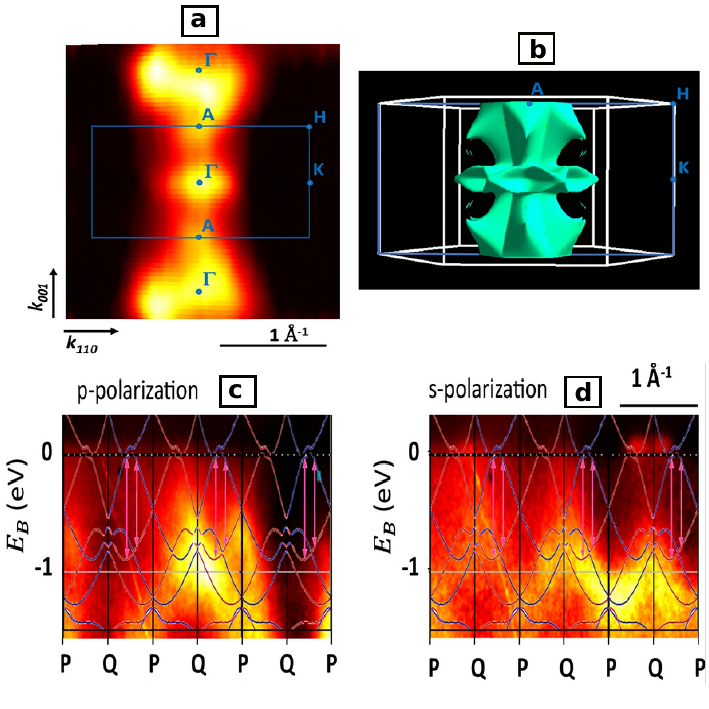} 
	\caption{a) Comparison of the SX-ARPES intensity at the Fermi energy obtained in the $\Gamma$-K-A plane with b) the computed three-dimensional Fermi surface. The corresponding Brillouin zone and high symmetry points are superimposed for clarity. The spin-integrated SX-ARPES intensity just below the Fermi energy is shown with the corresponding superimposed spin resolved band structure calculations: c) High symmetry P-Q path with p-polarized photons and d) P-Q path with s-polarized photons. These figures are adapted from Ref.\cite{reimers2024direct} under CC BY 4.0 International license. Copyright 2024, published by Springer Nature.  }
	\label{fig:your-label}
\end{figure}

The main altermagnetic characteristics were studied using angle-dependent SX-ARPES in thin epitaxial antiferromagnetic metallic films of CrSb\cite{reimers2024direct}, a predicted altermagnet\cite{vsmejkal2022emerging}. The measurements show a substantial spin split of about 0.6 eV below the Fermi level (see Fig.\ref{fig:your-label}), which is consistent with the expected result for altermagnets\cite{reimers2024direct}. CrSb has a high N\'eel temperature (703 K), which makes it a suitable altermagnet for applications at room temperature~\cite{zeng2024observation}. 

Exploring 2D materials, Liu et al.\cite{liu2024inverse} reported that FeS and FeSe are the thinnest 2D altermagnetic semiconductors with strong magnetic order, low exfoliation energy and good chemical stability, according to first-principles calculations. Both materials have A-type antiferromagnetic structures with large spin-splitting magnitudes (193 meV for FeS and 103 meV for FeSe) compared to other known altermagnets. The computation of the magnetic anisotropy energy (MAE) revealed that these materials are capable of preserving long-range magnetic order. 
     
A comprehensive computational analysis of the crystal and magnetic structures of MnF$_2$ was conducted by Bhowal et al.\cite{bhowal2022magnetic}, MnF$_2$ crystallizes in a centrosymmetric structure, the fluorine atoms around the manganese atom at the structure's center are rotated 90° around the z-axis compared to those surrounding the manganese atom at the structure's corner, leading to nonequivalent manganese sites although the presence of nonequivalent Mn sites raises questions about the exact symmetry classification.  DFT calculations, including LDA+U with a Hubbard potential (U$_\text{eff}$ = 5 eV), showed significant energy splitting even without SOC. The study also explored the role of magnetic octupoles, identifying their potential for tuning via alterations to the fluorine environment around the Mn sites without affecting the spin configuration. This tuning capability demonstrates that flipping the Mn spin arrangements reverses the sign of the octupole moment, highlighting magnetic octupoles as a natural-order parameter for NRSS antiferromagnetism. Magnetic Compton profile measurements were proposed as a tool to detect associated piezomagnetic and anti-piezomagnetic effects\cite{bhowal2022magnetic}. These findings suggest promising research directions, particularly the investigation of other magnetic materials with a centrosymmetric tetragonal rutile structure. The potential role of magnetic octupoles as order parameters in these systems warrants further detailed exploration.

\subsection{Contradicting reports on the magnetic ordering of RuO$_2$}
Although RuO$_2$ stands out as the most explored material for altermagnetism, however, its magnetic properties have been the subject of conflicting results with some suggesting Pauli paramagnetism\cite{ryden1970magnetic} in the past, while others indicate an antiferromagnetic configuration\cite{berlijn2017itinerant,zhu2019anomalous}.

The first principles calculation revealed NRSS up to 1.4 eV in RuO$_2$\cite{vsmejkal2022emerging} assuming a Hubbard-U of ~2 eV\cite{vsmejkal2020crystal}. However, assuming a zero U or a realistic value leads to a non-magnetic state as shown in the detailed study as a function of U for stoichiometric RuO$_2$ in Ref.~\cite{smolyanyuk2024fragility}. The study was also carried out for non-stoichiometric RuO$_2$, proposing that Ru vacancies could facilitate a magnetic state. This hypothesis is supported by the observation that hole-doped RuO$_2$ modeled by varying electron counts in DFT calculations can exhibit magnetic properties. Hence, the occurrence of altermagnetism depends on the sample's stoichiometry, with nonstoichiometric samples showing potential for magnetism due to Ru vacancies or hole doping.

The work of Smolyanuk et al.~\cite{smolyanyuk2024fragility} could provide a possible explanation for the large variability of experimental observations that could be attributed to different sample conditions, underscoring the need for more detailed characterization of RuO$_2$ stoichiometry in future studies. The magnetic characteristics of RuO$_2$ investigated in the study by combining a variety of cutting-edge spectroscopy techniques, including neutron diffraction and muon spin spectroscopy (${\mu}$SR), demonstrate that the magnetic signals previously found are probably the result of experimental errors such as multiple scattering at defects rather than intrinsic magnetic features of the material, therefore challenging prior assertions of magnetic order in RuO$_2$. The thoroughness of the methodology ensures high reliability in the results, particularly in ruling out magnetic order in RuO$_2$\cite{kessler2024absence,hiraishi2024nonmagnetic}. Furthermore, Hiraishi et al.\cite{
	hiraishi2024nonmagnetic} directly analyzes the band structures and spin polarization of single-crystal and thin-film rutile RuO$_2$ samples using ARPES and SX-ARPES.
The electronic structure of RuO$_2$ is reported to be reasonably consistent with non-magnetic conditions, as evidenced by the lack of k-dependent spin-splitting. 

Gomonay et al.\cite{gomonay2024structure} addressed RuO$_2$ by focusing on the peculiar magnon spectra and domain wall dynamics in the same material. The key contribution is a phenomenological theory to model altermagnets, revealing new magnetic properties such as lifted magnon degeneracy\cite{vsmejkal2023chiral,gohlke2023spurious} and anisotropic Walker breakdown. The micromagnetic framework is used to describe altermagnetic textures by extending traditional models used for ferromagnets and antiferromagnets. The model introduces additional order parameters specific to altermagnets. Employing Landau's phenomenological theory to explain the symmetry properties of altermagnets, where the sublattice spin symmetries give rise to the distinct magnetic behavior. In RuO$_2$ the model predicts an anisotropic spin stiffness, leading to variations in magnetization gradients within domain walls, distinguishing altermagnets from their antiferromagnetic counterparts. Altermagnetic domain walls are unique due to the anisotropy in the exchange stiffness. Unlike antiferromagnetic domain walls, altermagnets have domain walls with unequal sublattice magnetizations and widths. The gradient of magnetization creates a Ponderomotive force within the wall of the domain, allowing manipulation by magnetic force microscopy\cite{gomonay2024structure}. The analysis predicts a limited overall magnetization along the domain wall. Furthermore, investigating the motion of domain walls in altermagnets enabled the identification of an anisotropic Walker breakdown similar to ferromagnets but at higher velocities. Differences in sublattice stiffness lead to domain wall deformations during motion, reducing the domain wall velocity below the magnon speed. The analytical model relies on the Landau-Lifshitz equations describing the magnetization and N\'eel vector, adapted for altermagnets. The free energy functional includes both isotropic and anisotropic exchange stiffnesses to account for the altermagnetic behavior.

\subsection{Spin group theory description}
Landau theory is traditionally used for ferromagnets and antiferromagnets. The theoretical foundation laid by identifying a set of multipolar secondary order parameters bridges the gap between ideas about spin symmetries and the observable properties of these materials. McClarty et al. combined the Landau theory framework, emphasizing the importance of incorporating spin-space symmetry into the analysis\cite{mcclarty2024landau}. The thorough theoretical approach describes AM's behavior in a range of scenarios.
Recently, the approach used by Smejkal et al.\cite{vsmejkal2022beyond,vsmejkal2022emerging} to explain NRSS magnetic materials uses a description of magnetic group theory, where spin and space symmetries are decomposed. The transformation used as such $[C_i||C_j]$, the transformation left side of the double vertical line acts in the spin space, and the transformation right side of the double vertical line acts in real space. In case of RuO$_2$ the related spin group symmetry is $[C_2||C_{4z}t]$\cite{vsmejkal2022emerging}. Here twofold($C_2$) rotation in spin space and fourfold rotation($C_{4z}$) are merged with the translation (t) in the real space. If $r_s$ corresponds to a spin-only group containing the transformation solely in the spin space, and $R_s$ is the nontrivial spin group containing the transformation like $[C_i||C_j]$, however not containing elements of the explicit spin-only group, then the spin group can be expressed as the direct product $r_s\times R_s$\cite{litvin1977spin,litvin1974spin}. The explicit spin-only group is defined by $r_s$=$C_{\infty}$+$\bar{C}_2$$C_{\infty}$, here $C_{\infty}$ is the group containing all the spin space rotational transformations about the spins' common axis, and $\bar{C}_2$ is the two-fold rotation about an axis orthogonal to the spin, followed by reversal of spin space. Consequently, the symmetry operation $C_{\infty}$ makes spin a good quantum number. Hence, the band structure is accompanied by the split-band structure of the spin-up and spin-down channels. The three types of nontrivial spin Laue groups can describe different types of magnetic materials. The foremost kind is for ferromagnetic materials, where the group is represented by $R^I_s=[E||G]$, where G specifies the Laue group. $R^I_s$ defines the non-trivial spin group corresponding to conventional ferromagnetism (spin-split band structure with broken $\mathcal{T}$ symmetry). 
The second type of non-trivial spin group is $R^{II}_s=[E||G]+[C_2||G]$. The group $R^{II}_s$ also describes conventional antiferromagnetism and the related opposite spin sublattice symmetry is $[C_2||\bar{E}]$or$[C_2||t]$, where $\bar{E}$ is the real space inversion and t denotes translation.

$R^{III}_s=[E||H]+[C_2||AH]$ gives the nontrivial spin group that characterizes the altermagnets. Where $H$ is the Halving subgroup possessing half elements of the real space transformation of the nonmagnetic Laue group $G$, which includes the identity element of real space and $E$ stands for the spin space identity transformation. The $G-H=AH$ contains the residual half of the transformation, where $A$ is the real space rotation (proper or improper, symmorphic or non-symmorphic). Only real space transformations that exchange atoms between sublattices of the same spin are contained in $H$, and exclusive real space transformations that exchange atoms between sublattices of opposite spin are present in $G-H$. The non-trivial spin subgroup $[E||H]$ allows the symmetry in such a manner that spin sublattices are characterized by anisotropic spin densities. Similarly $[C_2||AH]$ is responsible for broken $\mathcal{T}$ symmetry and non-relativistic spin splitting of bands. Together, there are ten nontrivial spin Laue groups categorizing d-,g-,i-wave altermagnets\cite{vsmejkal2022emerging,vsmejkal2022beyond}.

\subsection{Elementary rules to identify altermagnets}
     A spin group theory developed by Litvin et al.\cite{litvin1974spin, litvin1980wreath} has been utilized in developing the selection rules for the altermagnets. The theoretical rules proposed by Smejkal et
al.\cite{vsmejkal2022emerging} to identify the altermagnets are as follows:

\begin{itemize}
	\item The number of magnetic atoms in a unit cell is even.
	\item The magnetic atoms in altermagnets are not related by inversion symmetry.
 \item Non-interconvertible local motif-pair spin anisotropy.
	\item The opposite-spin sublattices are connected by rotation of ($\pi/2$) or combined with translation or inversion symmetry. 
\end{itemize}

\hspace{0.5cm}Additionally, a computational tool has been developed by Smolyanyuk et al. to check whether a given material is an altermagnet or an antiferromagnet, which can read a Crystallographic Information File as input. 
There, users are asked to provide the spin configuration of the magnetic atom in U (up) or D (down) format. Moreover, if the atom is non-magnetic N needs to be given as input to finally get the result as an altermagnet being true or false\cite{smolyanyuk2024tool}.

\section{Conclusion}
To summarize, altermagnetism is an appealing, novel, and
distinct magnetic phase possessing zero net magnetization and strong time-reversal symmetry-breaking responses. It expands the traditional understanding of magnetism, which was previously limited to ferromagnetism and antiferromagnetism. The theoretical framework based on symmetry principles to classify and describe altermagnetism offers a clear distinction between ferromagnetic, antiferromagnetic, and altermagnetic phases.

Candidate altermagnetic materials range from insulators to metals as initially predicted by Smejkal et al.\cite{vsmejkal2022emerging} and were followed by several subsequent studies (see e.g. Refs.\cite{bhowal2022magnetic,chakraborty2024strain,lee2024broken,osumi2024observation}). Such predictions associated with state-of-the-art experiments are essential for guiding future research activities and exploring the technological application of altermagnetic materials in information technology by harnessing emerging spintronics phenomena, ultrafast photomagnetism, and thermoelectrics \cite{sukhachov2024thermoelectric,han2024observation,shao2021spin}. 
It is clear that there is a need for more experimental investigations to identify altermagnetic materials. 
The special properties of altermagnetism including strong spin coherence and high magnetic ordering temperatures make them promising candidates for next-generation devices. The use of multiple ARPES techniques ensures strong validation of the theoretical predictions. However, various studies only outline potential applications of altermagnetism, providing no detailed and complete guidance on the experimental methods required to realize these applications. More research is needed to bridge the gap between theoretical predictions and practical implementation. 


		\section*{Author contributions}
		\textbf{R. Tamang:} Formal analysis, Visualization, Validation, Literature review, Writing-original draft, writing-review \& editing.\\
		\textbf{Shivraj Gurung:} Formal analysis, Visualization, Validation, writing-review \& editing. \\ 
		\textbf{D. P. Rai:} Project management, Supervision, Resources, software, Formal analysis, Visualization, Validation, writing-review \& editing. \\ 
	    \textbf{Samy Brahimi}:Formal analysis, Visualization, Validation, writing-review \& editing.\\
		\textbf{Samir Lounis}: Supervision, Resources, Formal analysis, Visualization, Validation, writing-review \& editing.
		
		\begin{acknowledgments}
	DPR acknowledges the Science \& Engineering Research Board (SERB), New Delhi Govt. of India via File Number: SIR/2022/001150. \par RT acknowledges the University Grants Commission (UGC), India, for the Junior Research Fellowship (JRF), ID No. 231620066332.
 		\end{acknowledgments}
		

		\nocite{*}
		\bibliography{rsc.bib}

\begin{thebibliography}{89}%
\makeatletter
\providecommand \@ifxundefined [1]{%
 \@ifx{#1\undefined}
}%
\providecommand \@ifnum [1]{%
 \ifnum #1\expandafter \@firstoftwo
 \else \expandafter \@secondoftwo
 \fi
}%
\providecommand \@ifx [1]{%
 \ifx #1\expandafter \@firstoftwo
 \else \expandafter \@secondoftwo
 \fi
}%
\providecommand \natexlab [1]{#1}%
\providecommand \enquote  [1]{``#1''}%
\providecommand \bibnamefont  [1]{#1}%
\providecommand \bibfnamefont [1]{#1}%
\providecommand \citenamefont [1]{#1}%
\providecommand \href@noop [0]{\@secondoftwo}%
\providecommand \href [0]{\begingroup \@sanitize@url \@href}%
\providecommand \@href[1]{\@@startlink{#1}\@@href}%
\providecommand \@@href[1]{\endgroup#1\@@endlink}%
\providecommand \@sanitize@url [0]{\catcode `\\12\catcode `\$12\catcode `\&12\catcode `\#12\catcode `\^12\catcode `\_12\catcode `\%12\relax}%
\providecommand \@@startlink[1]{}%
\providecommand \@@endlink[0]{}%
\providecommand \url  [0]{\begingroup\@sanitize@url \@url }%
\providecommand \@url [1]{\endgroup\@href {#1}{\urlprefix }}%
\providecommand \urlprefix  [0]{URL }%
\providecommand \Eprint [0]{\href }%
\providecommand \doibase [0]{http://dx.doi.org/}%
\providecommand \selectlanguage [0]{\@gobble}%
\providecommand \bibinfo  [0]{\@secondoftwo}%
\providecommand \bibfield  [0]{\@secondoftwo}%
\providecommand \translation [1]{[#1]}%
\providecommand \BibitemOpen [0]{}%
\providecommand \bibitemStop [0]{}%
\providecommand \bibitemNoStop [0]{.\EOS\space}%
\providecommand \EOS [0]{\spacefactor3000\relax}%
\providecommand \BibitemShut  [1]{\csname bibitem#1\endcsname}%
\let\auto@bib@innerbib\@empty
\bibitem [{\citenamefont {{\v{S}}mejkal}\ \emph {et~al.}(2022{\natexlab{a}})\citenamefont {{\v{S}}mejkal}, \citenamefont {Sinova},\ and\ \citenamefont {Jungwirth}}]{vsmejkal2022beyond}%
  \BibitemOpen
  \bibfield  {author} {\bibinfo {author} {\bibfnamefont {L.}~\bibnamefont {{\v{S}}mejkal}}, \bibinfo {author} {\bibfnamefont {J.}~\bibnamefont {Sinova}}, \ and\ \bibinfo {author} {\bibfnamefont {T.}~\bibnamefont {Jungwirth}},\ }\href@noop {} {\bibfield  {journal} {\bibinfo  {journal} {Physical Review X}\ }\textbf {\bibinfo {volume} {12}},\ \bibinfo {pages} {031042} (\bibinfo {year} {2022}{\natexlab{a}})}\BibitemShut {NoStop}%
\bibitem [{\citenamefont {N{\'e}el}(1953)}]{neel1953some}%
  \BibitemOpen
  \bibfield  {author} {\bibinfo {author} {\bibfnamefont {L.}~\bibnamefont {N{\'e}el}},\ }\href@noop {} {\bibfield  {journal} {\bibinfo  {journal} {Reviews of Modern Physics}\ }\textbf {\bibinfo {volume} {25}},\ \bibinfo {pages} {58} (\bibinfo {year} {1953})}\BibitemShut {NoStop}%
\bibitem [{\citenamefont {{\v{S}}mejkal}\ \emph {et~al.}(2022{\natexlab{b}})\citenamefont {{\v{S}}mejkal}, \citenamefont {Sinova},\ and\ \citenamefont {Jungwirth}}]{vsmejkal2022emerging}%
  \BibitemOpen
  \bibfield  {author} {\bibinfo {author} {\bibfnamefont {L.}~\bibnamefont {{\v{S}}mejkal}}, \bibinfo {author} {\bibfnamefont {J.}~\bibnamefont {Sinova}}, \ and\ \bibinfo {author} {\bibfnamefont {T.}~\bibnamefont {Jungwirth}},\ }\href@noop {} {\bibfield  {journal} {\bibinfo  {journal} {Physical Review X}\ }\textbf {\bibinfo {volume} {12}},\ \bibinfo {pages} {040501} (\bibinfo {year} {2022}{\natexlab{b}})}\BibitemShut {NoStop}%
\bibitem [{\citenamefont {Shao}\ \emph {et~al.}(2021)\citenamefont {Shao}, \citenamefont {Zhang}, \citenamefont {Li}, \citenamefont {Eom},\ and\ \citenamefont {Tsymbal}}]{shao2021spin}%
  \BibitemOpen
  \bibfield  {author} {\bibinfo {author} {\bibfnamefont {D.-F.}\ \bibnamefont {Shao}}, \bibinfo {author} {\bibfnamefont {S.-H.}\ \bibnamefont {Zhang}}, \bibinfo {author} {\bibfnamefont {M.}~\bibnamefont {Li}}, \bibinfo {author} {\bibfnamefont {C.-B.}\ \bibnamefont {Eom}}, \ and\ \bibinfo {author} {\bibfnamefont {E.~Y.}\ \bibnamefont {Tsymbal}},\ }\href@noop {} {\bibfield  {journal} {\bibinfo  {journal} {Nature Communications}\ }\textbf {\bibinfo {volume} {12}},\ \bibinfo {pages} {7061} (\bibinfo {year} {2021})}\BibitemShut {NoStop}%
\bibitem [{\citenamefont {Gomonay}\ \emph {et~al.}(2024)\citenamefont {Gomonay}, \citenamefont {Kravchuk}, \citenamefont {Jaeschke-Ubiergo}, \citenamefont {Yershov}, \citenamefont {Jungwirth}, \citenamefont {{\v{S}}mejkal}, \citenamefont {Brink},\ and\ \citenamefont {Sinova}}]{gomonay2024structure}%
  \BibitemOpen
  \bibfield  {author} {\bibinfo {author} {\bibfnamefont {O.}~\bibnamefont {Gomonay}}, \bibinfo {author} {\bibfnamefont {V.}~\bibnamefont {Kravchuk}}, \bibinfo {author} {\bibfnamefont {R.}~\bibnamefont {Jaeschke-Ubiergo}}, \bibinfo {author} {\bibfnamefont {K.}~\bibnamefont {Yershov}}, \bibinfo {author} {\bibfnamefont {T.}~\bibnamefont {Jungwirth}}, \bibinfo {author} {\bibfnamefont {L.}~\bibnamefont {{\v{S}}mejkal}}, \bibinfo {author} {\bibfnamefont {J.~v.~d.}\ \bibnamefont {Brink}}, \ and\ \bibinfo {author} {\bibfnamefont {J.}~\bibnamefont {Sinova}},\ }\href@noop {} {\bibfield  {journal} {\bibinfo  {journal} {npj Spintronics}\ }\textbf {\bibinfo {volume} {2}},\ \bibinfo {pages} {35} (\bibinfo {year} {2024})}\BibitemShut {NoStop}%
\bibitem [{\citenamefont {Jungwirth}\ \emph {et~al.}(2024)\citenamefont {Jungwirth}, \citenamefont {Fernandes}, \citenamefont {Sinova},\ and\ \citenamefont {Smejkal}}]{jungwirth2024altermagnets}%
  \BibitemOpen
  \bibfield  {author} {\bibinfo {author} {\bibfnamefont {T.}~\bibnamefont {Jungwirth}}, \bibinfo {author} {\bibfnamefont {R.~M.}\ \bibnamefont {Fernandes}}, \bibinfo {author} {\bibfnamefont {J.}~\bibnamefont {Sinova}}, \ and\ \bibinfo {author} {\bibfnamefont {L.}~\bibnamefont {Smejkal}},\ }\href@noop {} {\bibfield  {journal} {\bibinfo  {journal} {arXiv preprint arXiv:2409.10034}\ } (\bibinfo {year} {2024})}\BibitemShut {NoStop}%
\bibitem [{\citenamefont {Wu}\ \emph {et~al.}(2007)\citenamefont {Wu}, \citenamefont {Sun}, \citenamefont {Fradkin},\ and\ \citenamefont {Zhang}}]{wu2007fermi}%
  \BibitemOpen
  \bibfield  {author} {\bibinfo {author} {\bibfnamefont {C.}~\bibnamefont {Wu}}, \bibinfo {author} {\bibfnamefont {K.}~\bibnamefont {Sun}}, \bibinfo {author} {\bibfnamefont {E.}~\bibnamefont {Fradkin}}, \ and\ \bibinfo {author} {\bibfnamefont {S.-C.}\ \bibnamefont {Zhang}},\ }\href@noop {} {\bibfield  {journal} {\bibinfo  {journal} {Physical Review B—Condensed Matter and Materials Physics}\ }\textbf {\bibinfo {volume} {75}},\ \bibinfo {pages} {115103} (\bibinfo {year} {2007})}\BibitemShut {NoStop}%
\bibitem [{\citenamefont {Pomeranchuk}\ \emph {et~al.}(1958)\citenamefont {Pomeranchuk} \emph {et~al.}}]{pomeranchuk1958stability}%
  \BibitemOpen
  \bibfield  {author} {\bibinfo {author} {\bibfnamefont {I.~I.}\ \bibnamefont {Pomeranchuk}} \emph {et~al.},\ }\href@noop {} {\bibfield  {journal} {\bibinfo  {journal} {Sov. Phys. JETP}\ }\textbf {\bibinfo {volume} {8}},\ \bibinfo {pages} {361} (\bibinfo {year} {1958})}\BibitemShut {NoStop}%
\bibitem [{\citenamefont {Mazin}\ and\ \citenamefont {Editors}(2022)}]{mazin2022altermagnetism}%
  \BibitemOpen
  \bibfield  {author} {\bibinfo {author} {\bibfnamefont {I.}~\bibnamefont {Mazin}}\ and\ \bibinfo {author} {\bibfnamefont {P.}~\bibnamefont {Editors}},\ }\href@noop {} {\enquote {\bibinfo {title} {Altermagnetism—a new punch line of fundamental magnetism},}\ } (\bibinfo {year} {2022})\BibitemShut {NoStop}%
\bibitem [{\citenamefont {Cheong}\ and\ \citenamefont {Huang}(2024)}]{cheong2024altermagnetism}%
  \BibitemOpen
  \bibfield  {author} {\bibinfo {author} {\bibfnamefont {S.-W.}\ \bibnamefont {Cheong}}\ and\ \bibinfo {author} {\bibfnamefont {F.-T.}\ \bibnamefont {Huang}},\ }\href@noop {} {\bibfield  {journal} {\bibinfo  {journal} {npj Quantum Materials}\ }\textbf {\bibinfo {volume} {9}},\ \bibinfo {pages} {13} (\bibinfo {year} {2024})}\BibitemShut {NoStop}%
\bibitem [{\citenamefont {Brown}\ and\ \citenamefont {Chatterji}(2006)}]{brown2006neutron}%
  \BibitemOpen
  \bibfield  {author} {\bibinfo {author} {\bibfnamefont {P.}~\bibnamefont {Brown}}\ and\ \bibinfo {author} {\bibfnamefont {T.}~\bibnamefont {Chatterji}},\ }\href@noop {} {\bibfield  {journal} {\bibinfo  {journal} {Journal of Physics: Condensed Matter}\ }\textbf {\bibinfo {volume} {18}},\ \bibinfo {pages} {10085} (\bibinfo {year} {2006})}\BibitemShut {NoStop}%
\bibitem [{\citenamefont {Gonz{\'a}lez-Hern{\'a}ndez}\ \emph {et~al.}(2021)\citenamefont {Gonz{\'a}lez-Hern{\'a}ndez}, \citenamefont {{\v{S}}mejkal}, \citenamefont {V{\`y}born{\`y}}, \citenamefont {Yahagi}, \citenamefont {Sinova}, \citenamefont {Jungwirth},\ and\ \citenamefont {{\v{Z}}elezn{\`y}}}]{gonzalez2021efficient}%
  \BibitemOpen
  \bibfield  {author} {\bibinfo {author} {\bibfnamefont {R.}~\bibnamefont {Gonz{\'a}lez-Hern{\'a}ndez}}, \bibinfo {author} {\bibfnamefont {L.}~\bibnamefont {{\v{S}}mejkal}}, \bibinfo {author} {\bibfnamefont {K.}~\bibnamefont {V{\`y}born{\`y}}}, \bibinfo {author} {\bibfnamefont {Y.}~\bibnamefont {Yahagi}}, \bibinfo {author} {\bibfnamefont {J.}~\bibnamefont {Sinova}}, \bibinfo {author} {\bibfnamefont {T.}~\bibnamefont {Jungwirth}}, \ and\ \bibinfo {author} {\bibfnamefont {J.}~\bibnamefont {{\v{Z}}elezn{\`y}}},\ }\href@noop {} {\bibfield  {journal} {\bibinfo  {journal} {Physical Review Letters}\ }\textbf {\bibinfo {volume} {126}},\ \bibinfo {pages} {127701} (\bibinfo {year} {2021})}\BibitemShut {NoStop}%
\bibitem [{\citenamefont {Bose}\ \emph {et~al.}(2022)\citenamefont {Bose}, \citenamefont {Schreiber}, \citenamefont {Jain}, \citenamefont {Shao}, \citenamefont {Nair}, \citenamefont {Sun}, \citenamefont {Zhang}, \citenamefont {Muller}, \citenamefont {Tsymbal}, \citenamefont {Schlom} \emph {et~al.}}]{bose2022tilted}%
  \BibitemOpen
  \bibfield  {author} {\bibinfo {author} {\bibfnamefont {A.}~\bibnamefont {Bose}}, \bibinfo {author} {\bibfnamefont {N.~J.}\ \bibnamefont {Schreiber}}, \bibinfo {author} {\bibfnamefont {R.}~\bibnamefont {Jain}}, \bibinfo {author} {\bibfnamefont {D.-F.}\ \bibnamefont {Shao}}, \bibinfo {author} {\bibfnamefont {H.~P.}\ \bibnamefont {Nair}}, \bibinfo {author} {\bibfnamefont {J.}~\bibnamefont {Sun}}, \bibinfo {author} {\bibfnamefont {X.~S.}\ \bibnamefont {Zhang}}, \bibinfo {author} {\bibfnamefont {D.~A.}\ \bibnamefont {Muller}}, \bibinfo {author} {\bibfnamefont {E.~Y.}\ \bibnamefont {Tsymbal}}, \bibinfo {author} {\bibfnamefont {D.~G.}\ \bibnamefont {Schlom}},  \emph {et~al.},\ }\href@noop {} {\bibfield  {journal} {\bibinfo  {journal} {Nature Electronics}\ }\textbf {\bibinfo {volume} {5}},\ \bibinfo {pages} {267} (\bibinfo {year} {2022})}\BibitemShut {NoStop}%
\bibitem [{\citenamefont {Bai}\ \emph {et~al.}(2022)\citenamefont {Bai}, \citenamefont {Han}, \citenamefont {Feng}, \citenamefont {Zhou}, \citenamefont {Su}, \citenamefont {Wang}, \citenamefont {Liao}, \citenamefont {Zhu}, \citenamefont {Chen}, \citenamefont {Pan} \emph {et~al.}}]{bai2022observation}%
  \BibitemOpen
  \bibfield  {author} {\bibinfo {author} {\bibfnamefont {H.}~\bibnamefont {Bai}}, \bibinfo {author} {\bibfnamefont {L.}~\bibnamefont {Han}}, \bibinfo {author} {\bibfnamefont {X.}~\bibnamefont {Feng}}, \bibinfo {author} {\bibfnamefont {Y.}~\bibnamefont {Zhou}}, \bibinfo {author} {\bibfnamefont {R.}~\bibnamefont {Su}}, \bibinfo {author} {\bibfnamefont {Q.}~\bibnamefont {Wang}}, \bibinfo {author} {\bibfnamefont {L.}~\bibnamefont {Liao}}, \bibinfo {author} {\bibfnamefont {W.}~\bibnamefont {Zhu}}, \bibinfo {author} {\bibfnamefont {X.}~\bibnamefont {Chen}}, \bibinfo {author} {\bibfnamefont {F.}~\bibnamefont {Pan}},  \emph {et~al.},\ }\href@noop {} {\bibfield  {journal} {\bibinfo  {journal} {Physical Review Letters}\ }\textbf {\bibinfo {volume} {128}},\ \bibinfo {pages} {197202} (\bibinfo {year} {2022})}\BibitemShut {NoStop}%
\bibitem [{\citenamefont {Karube}\ \emph {et~al.}(2022)\citenamefont {Karube}, \citenamefont {Tanaka}, \citenamefont {Sugawara}, \citenamefont {Kadoguchi}, \citenamefont {Kohda},\ and\ \citenamefont {Nitta}}]{karube2022observation}%
  \BibitemOpen
  \bibfield  {author} {\bibinfo {author} {\bibfnamefont {S.}~\bibnamefont {Karube}}, \bibinfo {author} {\bibfnamefont {T.}~\bibnamefont {Tanaka}}, \bibinfo {author} {\bibfnamefont {D.}~\bibnamefont {Sugawara}}, \bibinfo {author} {\bibfnamefont {N.}~\bibnamefont {Kadoguchi}}, \bibinfo {author} {\bibfnamefont {M.}~\bibnamefont {Kohda}}, \ and\ \bibinfo {author} {\bibfnamefont {J.}~\bibnamefont {Nitta}},\ }\href@noop {} {\bibfield  {journal} {\bibinfo  {journal} {Physical review letters}\ }\textbf {\bibinfo {volume} {129}},\ \bibinfo {pages} {137201} (\bibinfo {year} {2022})}\BibitemShut {NoStop}%
\bibitem [{\citenamefont {{\v{S}}mejkal}\ \emph {et~al.}(2021)\citenamefont {{\v{S}}mejkal}, \citenamefont {Hellenes}, \citenamefont {Gonz{\'a}lez-Hern{\'a}ndez}, \citenamefont {Sinova},\ and\ \citenamefont {Jungwirth}}]{vsmejkal2021giant}%
  \BibitemOpen
  \bibfield  {author} {\bibinfo {author} {\bibfnamefont {L.}~\bibnamefont {{\v{S}}mejkal}}, \bibinfo {author} {\bibfnamefont {A.~B.}\ \bibnamefont {Hellenes}}, \bibinfo {author} {\bibfnamefont {R.}~\bibnamefont {Gonz{\'a}lez-Hern{\'a}ndez}}, \bibinfo {author} {\bibfnamefont {J.}~\bibnamefont {Sinova}}, \ and\ \bibinfo {author} {\bibfnamefont {T.}~\bibnamefont {Jungwirth}},\ }\href@noop {} {\bibfield  {journal} {\bibinfo  {journal} {arXiv preprint arXiv:2103.12664}\ } (\bibinfo {year} {2021})}\BibitemShut {NoStop}%
\bibitem [{\citenamefont {Baltz}\ \emph {et~al.}(2024)\citenamefont {Baltz}, \citenamefont {Hoffmann}, \citenamefont {Emori}, \citenamefont {Shao},\ and\ \citenamefont {Jungwirth}}]{baltz2024emerging}%
  \BibitemOpen
  \bibfield  {author} {\bibinfo {author} {\bibfnamefont {V.}~\bibnamefont {Baltz}}, \bibinfo {author} {\bibfnamefont {A.}~\bibnamefont {Hoffmann}}, \bibinfo {author} {\bibfnamefont {S.}~\bibnamefont {Emori}}, \bibinfo {author} {\bibfnamefont {D.-F.}\ \bibnamefont {Shao}}, \ and\ \bibinfo {author} {\bibfnamefont {T.}~\bibnamefont {Jungwirth}},\ }\href@noop {} {\bibfield  {journal} {\bibinfo  {journal} {APL Materials}\ }\textbf {\bibinfo {volume} {12}} (\bibinfo {year} {2024})}\BibitemShut {NoStop}%
\bibitem [{\citenamefont {Qiu}\ \emph {et~al.}(2023)\citenamefont {Qiu}, \citenamefont {Seifert}, \citenamefont {Huang}, \citenamefont {Zhou}, \citenamefont {Ka{\v{s}}par}, \citenamefont {Zhang}, \citenamefont {Wu}, \citenamefont {Fan}, \citenamefont {Zhang}, \citenamefont {Wu} \emph {et~al.}}]{qiu2023terahertz}%
  \BibitemOpen
  \bibfield  {author} {\bibinfo {author} {\bibfnamefont {H.}~\bibnamefont {Qiu}}, \bibinfo {author} {\bibfnamefont {T.~S.}\ \bibnamefont {Seifert}}, \bibinfo {author} {\bibfnamefont {L.}~\bibnamefont {Huang}}, \bibinfo {author} {\bibfnamefont {Y.}~\bibnamefont {Zhou}}, \bibinfo {author} {\bibfnamefont {Z.}~\bibnamefont {Ka{\v{s}}par}}, \bibinfo {author} {\bibfnamefont {C.}~\bibnamefont {Zhang}}, \bibinfo {author} {\bibfnamefont {J.}~\bibnamefont {Wu}}, \bibinfo {author} {\bibfnamefont {K.}~\bibnamefont {Fan}}, \bibinfo {author} {\bibfnamefont {Q.}~\bibnamefont {Zhang}}, \bibinfo {author} {\bibfnamefont {D.}~\bibnamefont {Wu}},  \emph {et~al.},\ }\href@noop {} {\bibfield  {journal} {\bibinfo  {journal} {Advanced Science}\ }\textbf {\bibinfo {volume} {10}},\ \bibinfo {pages} {2300512} (\bibinfo {year} {2023})}\BibitemShut {NoStop}%
\bibitem [{\citenamefont {Reichlova}\ \emph {et~al.}(2024)\citenamefont {Reichlova}, \citenamefont {Kriegner}, \citenamefont {Mook}, \citenamefont {Althammer},\ and\ \citenamefont {Thomas}}]{reichlova2024role}%
  \BibitemOpen
  \bibfield  {author} {\bibinfo {author} {\bibfnamefont {H.}~\bibnamefont {Reichlova}}, \bibinfo {author} {\bibfnamefont {D.}~\bibnamefont {Kriegner}}, \bibinfo {author} {\bibfnamefont {A.}~\bibnamefont {Mook}}, \bibinfo {author} {\bibfnamefont {M.}~\bibnamefont {Althammer}}, \ and\ \bibinfo {author} {\bibfnamefont {A.}~\bibnamefont {Thomas}},\ }\href@noop {} {\bibfield  {journal} {\bibinfo  {journal} {APL Materials}\ }\textbf {\bibinfo {volume} {12}} (\bibinfo {year} {2024})}\BibitemShut {NoStop}%
\bibitem [{\citenamefont {Zhu}\ \emph {et~al.}(2023)\citenamefont {Zhu}, \citenamefont {Zhuang}, \citenamefont {Wu},\ and\ \citenamefont {Yan}}]{zhu2023topological}%
  \BibitemOpen
  \bibfield  {author} {\bibinfo {author} {\bibfnamefont {D.}~\bibnamefont {Zhu}}, \bibinfo {author} {\bibfnamefont {Z.-Y.}\ \bibnamefont {Zhuang}}, \bibinfo {author} {\bibfnamefont {Z.}~\bibnamefont {Wu}}, \ and\ \bibinfo {author} {\bibfnamefont {Z.}~\bibnamefont {Yan}},\ }\href@noop {} {\bibfield  {journal} {\bibinfo  {journal} {Physical Review B}\ }\textbf {\bibinfo {volume} {108}},\ \bibinfo {pages} {184505} (\bibinfo {year} {2023})}\BibitemShut {NoStop}%
\bibitem [{\citenamefont {Li}\ \emph {et~al.}(2024)\citenamefont {Li}, \citenamefont {Liu},\ and\ \citenamefont {Liu}}]{li2024creation}%
  \BibitemOpen
  \bibfield  {author} {\bibinfo {author} {\bibfnamefont {Y.-X.}\ \bibnamefont {Li}}, \bibinfo {author} {\bibfnamefont {Y.}~\bibnamefont {Liu}}, \ and\ \bibinfo {author} {\bibfnamefont {C.-C.}\ \bibnamefont {Liu}},\ }\href@noop {} {\bibfield  {journal} {\bibinfo  {journal} {Physical Review B}\ }\textbf {\bibinfo {volume} {109}},\ \bibinfo {pages} {L201109} (\bibinfo {year} {2024})}\BibitemShut {NoStop}%
\bibitem [{\citenamefont {Kang}\ \emph {et~al.}(2015)\citenamefont {Kang}, \citenamefont {Zhang}, \citenamefont {Wang}, \citenamefont {Klein}, \citenamefont {Chappert}, \citenamefont {Ravelosona}, \citenamefont {Wang}, \citenamefont {Zhang},\ and\ \citenamefont {Zhao}}]{kang2015spintronics}%
  \BibitemOpen
  \bibfield  {author} {\bibinfo {author} {\bibfnamefont {W.}~\bibnamefont {Kang}}, \bibinfo {author} {\bibfnamefont {Y.}~\bibnamefont {Zhang}}, \bibinfo {author} {\bibfnamefont {Z.}~\bibnamefont {Wang}}, \bibinfo {author} {\bibfnamefont {J.-O.}\ \bibnamefont {Klein}}, \bibinfo {author} {\bibfnamefont {C.}~\bibnamefont {Chappert}}, \bibinfo {author} {\bibfnamefont {D.}~\bibnamefont {Ravelosona}}, \bibinfo {author} {\bibfnamefont {G.}~\bibnamefont {Wang}}, \bibinfo {author} {\bibfnamefont {Y.}~\bibnamefont {Zhang}}, \ and\ \bibinfo {author} {\bibfnamefont {W.}~\bibnamefont {Zhao}},\ }\href@noop {} {\bibfield  {journal} {\bibinfo  {journal} {ACM Journal on Emerging Technologies in Computing Systems (JETC)}\ }\textbf {\bibinfo {volume} {12}},\ \bibinfo {pages} {1} (\bibinfo {year} {2015})}\BibitemShut {NoStop}%
\bibitem [{\citenamefont {Joshi}(2016)}]{joshi2016spintronics}%
  \BibitemOpen
  \bibfield  {author} {\bibinfo {author} {\bibfnamefont {V.~K.}\ \bibnamefont {Joshi}},\ }\href@noop {} {\bibfield  {journal} {\bibinfo  {journal} {Engineering science and technology, an international journal}\ }\textbf {\bibinfo {volume} {19}},\ \bibinfo {pages} {1503} (\bibinfo {year} {2016})}\BibitemShut {NoStop}%
\bibitem [{\citenamefont {McClarty}\ and\ \citenamefont {Rau}(2024)}]{mcclarty2024landau}%
  \BibitemOpen
  \bibfield  {author} {\bibinfo {author} {\bibfnamefont {P.~A.}\ \bibnamefont {McClarty}}\ and\ \bibinfo {author} {\bibfnamefont {J.~G.}\ \bibnamefont {Rau}},\ }\href@noop {} {\bibfield  {journal} {\bibinfo  {journal} {Physical Review Letters}\ }\textbf {\bibinfo {volume} {132}},\ \bibinfo {pages} {176702} (\bibinfo {year} {2024})}\BibitemShut {NoStop}%
\bibitem [{\citenamefont {Bai}\ \emph {et~al.}(2024)\citenamefont {Bai}, \citenamefont {Feng}, \citenamefont {Liu}, \citenamefont {{\v{S}}mejkal}, \citenamefont {Mokrousov},\ and\ \citenamefont {Yao}}]{bai2024altermagnetism}%
  \BibitemOpen
  \bibfield  {author} {\bibinfo {author} {\bibfnamefont {L.}~\bibnamefont {Bai}}, \bibinfo {author} {\bibfnamefont {W.}~\bibnamefont {Feng}}, \bibinfo {author} {\bibfnamefont {S.}~\bibnamefont {Liu}}, \bibinfo {author} {\bibfnamefont {L.}~\bibnamefont {{\v{S}}mejkal}}, \bibinfo {author} {\bibfnamefont {Y.}~\bibnamefont {Mokrousov}}, \ and\ \bibinfo {author} {\bibfnamefont {Y.}~\bibnamefont {Yao}},\ }\href@noop {} {\bibfield  {journal} {\bibinfo  {journal} {Advanced Functional Materials}\ ,\ \bibinfo {pages} {2409327}} (\bibinfo {year} {2024})}\BibitemShut {NoStop}%
\bibitem [{\citenamefont {Gao}\ \emph {et~al.}(2023)\citenamefont {Gao}, \citenamefont {Qu}, \citenamefont {Zeng}, \citenamefont {Wen}, \citenamefont {Sun}, \citenamefont {Guo},\ and\ \citenamefont {Lu}}]{gao2023ai}%
  \BibitemOpen
  \bibfield  {author} {\bibinfo {author} {\bibfnamefont {Z.-F.}\ \bibnamefont {Gao}}, \bibinfo {author} {\bibfnamefont {S.}~\bibnamefont {Qu}}, \bibinfo {author} {\bibfnamefont {B.}~\bibnamefont {Zeng}}, \bibinfo {author} {\bibfnamefont {J.-R.}\ \bibnamefont {Wen}}, \bibinfo {author} {\bibfnamefont {H.}~\bibnamefont {Sun}}, \bibinfo {author} {\bibfnamefont {P.}~\bibnamefont {Guo}}, \ and\ \bibinfo {author} {\bibfnamefont {Z.-Y.}\ \bibnamefont {Lu}},\ }\href@noop {} {\bibfield  {journal} {\bibinfo  {journal} {arXiv preprint arXiv:2311.04418}\ } (\bibinfo {year} {2023})}\BibitemShut {NoStop}%
\bibitem [{\citenamefont {He}\ \emph {et~al.}(2023)\citenamefont {He}, \citenamefont {Wang}, \citenamefont {Luo}, \citenamefont {Zeng}, \citenamefont {Chen},\ and\ \citenamefont {Tang}}]{he2023nonrelativistic}%
  \BibitemOpen
  \bibfield  {author} {\bibinfo {author} {\bibfnamefont {R.}~\bibnamefont {He}}, \bibinfo {author} {\bibfnamefont {D.}~\bibnamefont {Wang}}, \bibinfo {author} {\bibfnamefont {N.}~\bibnamefont {Luo}}, \bibinfo {author} {\bibfnamefont {J.}~\bibnamefont {Zeng}}, \bibinfo {author} {\bibfnamefont {K.-Q.}\ \bibnamefont {Chen}}, \ and\ \bibinfo {author} {\bibfnamefont {L.-M.}\ \bibnamefont {Tang}},\ }\href@noop {} {\bibfield  {journal} {\bibinfo  {journal} {Physical Review Letters}\ }\textbf {\bibinfo {volume} {130}},\ \bibinfo {pages} {046401} (\bibinfo {year} {2023})}\BibitemShut {NoStop}%
\bibitem [{\citenamefont {Pan}\ \emph {et~al.}(2024)\citenamefont {Pan}, \citenamefont {Zhou}, \citenamefont {Lyu}, \citenamefont {Xiao}, \citenamefont {Yang},\ and\ \citenamefont {Sun}}]{pan2024general}%
  \BibitemOpen
  \bibfield  {author} {\bibinfo {author} {\bibfnamefont {B.}~\bibnamefont {Pan}}, \bibinfo {author} {\bibfnamefont {P.}~\bibnamefont {Zhou}}, \bibinfo {author} {\bibfnamefont {P.}~\bibnamefont {Lyu}}, \bibinfo {author} {\bibfnamefont {H.}~\bibnamefont {Xiao}}, \bibinfo {author} {\bibfnamefont {X.}~\bibnamefont {Yang}}, \ and\ \bibinfo {author} {\bibfnamefont {L.}~\bibnamefont {Sun}},\ }\href@noop {} {\bibfield  {journal} {\bibinfo  {journal} {Physical Review Letters}\ }\textbf {\bibinfo {volume} {133}},\ \bibinfo {pages} {166701} (\bibinfo {year} {2024})}\BibitemShut {NoStop}%
\bibitem [{\citenamefont {Chakraborty}\ \emph {et~al.}(2024)\citenamefont {Chakraborty}, \citenamefont {Gonz{\'a}lez~Hern{\'a}ndez}, \citenamefont {{\v{S}}mejkal},\ and\ \citenamefont {Sinova}}]{chakraborty2024strain}%
  \BibitemOpen
  \bibfield  {author} {\bibinfo {author} {\bibfnamefont {A.}~\bibnamefont {Chakraborty}}, \bibinfo {author} {\bibfnamefont {R.}~\bibnamefont {Gonz{\'a}lez~Hern{\'a}ndez}}, \bibinfo {author} {\bibfnamefont {L.}~\bibnamefont {{\v{S}}mejkal}}, \ and\ \bibinfo {author} {\bibfnamefont {J.}~\bibnamefont {Sinova}},\ }\href@noop {} {\bibfield  {journal} {\bibinfo  {journal} {Physical Review B}\ }\textbf {\bibinfo {volume} {109}},\ \bibinfo {pages} {144421} (\bibinfo {year} {2024})}\BibitemShut {NoStop}%
\bibitem [{\citenamefont {Rashba}(1960)}]{rashba1960spin}%
  \BibitemOpen
  \bibfield  {author} {\bibinfo {author} {\bibfnamefont {E.~I.}\ \bibnamefont {Rashba}},\ }\href@noop {} {\bibfield  {journal} {\bibinfo  {journal} {Sov. Phys. Solid State}\ }\textbf {\bibinfo {volume} {2}},\ \bibinfo {pages} {1109} (\bibinfo {year} {1960})}\BibitemShut {NoStop}%
\bibitem [{\citenamefont {Dresselhaus}(1955)}]{dresselhaus1955spin}%
  \BibitemOpen
  \bibfield  {author} {\bibinfo {author} {\bibfnamefont {G.}~\bibnamefont {Dresselhaus}},\ }\href@noop {} {\bibfield  {journal} {\bibinfo  {journal} {Physical Review}\ }\textbf {\bibinfo {volume} {100}},\ \bibinfo {pages} {580} (\bibinfo {year} {1955})}\BibitemShut {NoStop}%
\bibitem [{\citenamefont {Yuan}\ \emph {et~al.}(2020)\citenamefont {Yuan}, \citenamefont {Wang}, \citenamefont {Luo}, \citenamefont {Rashba},\ and\ \citenamefont {Zunger}}]{yuan2020giant}%
  \BibitemOpen
  \bibfield  {author} {\bibinfo {author} {\bibfnamefont {L.-D.}\ \bibnamefont {Yuan}}, \bibinfo {author} {\bibfnamefont {Z.}~\bibnamefont {Wang}}, \bibinfo {author} {\bibfnamefont {J.-W.}\ \bibnamefont {Luo}}, \bibinfo {author} {\bibfnamefont {E.~I.}\ \bibnamefont {Rashba}}, \ and\ \bibinfo {author} {\bibfnamefont {A.}~\bibnamefont {Zunger}},\ }\href@noop {} {\bibfield  {journal} {\bibinfo  {journal} {Physical Review B}\ }\textbf {\bibinfo {volume} {102}},\ \bibinfo {pages} {014422} (\bibinfo {year} {2020})}\BibitemShut {NoStop}%
\bibitem [{\citenamefont {Sattigeri}\ \emph {et~al.}(2023)\citenamefont {Sattigeri}, \citenamefont {Cuono},\ and\ \citenamefont {Autieri}}]{sattigeri2023altermagnetic}%
  \BibitemOpen
  \bibfield  {author} {\bibinfo {author} {\bibfnamefont {R.~M.}\ \bibnamefont {Sattigeri}}, \bibinfo {author} {\bibfnamefont {G.}~\bibnamefont {Cuono}}, \ and\ \bibinfo {author} {\bibfnamefont {C.}~\bibnamefont {Autieri}},\ }\href@noop {} {\bibfield  {journal} {\bibinfo  {journal} {Nanoscale}\ }\textbf {\bibinfo {volume} {15}},\ \bibinfo {pages} {16998} (\bibinfo {year} {2023})}\BibitemShut {NoStop}%
\bibitem [{\citenamefont {{\v{S}}mejkal}\ \emph {et~al.}(2020)\citenamefont {{\v{S}}mejkal}, \citenamefont {Gonz{\'a}lez-Hern{\'a}ndez}, \citenamefont {Jungwirth},\ and\ \citenamefont {Sinova}}]{vsmejkal2020crystal}%
  \BibitemOpen
  \bibfield  {author} {\bibinfo {author} {\bibfnamefont {L.}~\bibnamefont {{\v{S}}mejkal}}, \bibinfo {author} {\bibfnamefont {R.}~\bibnamefont {Gonz{\'a}lez-Hern{\'a}ndez}}, \bibinfo {author} {\bibfnamefont {T.}~\bibnamefont {Jungwirth}}, \ and\ \bibinfo {author} {\bibfnamefont {J.}~\bibnamefont {Sinova}},\ }\href@noop {} {\bibfield  {journal} {\bibinfo  {journal} {Science advances}\ }\textbf {\bibinfo {volume} {6}},\ \bibinfo {pages} {eaaz8809} (\bibinfo {year} {2020})}\BibitemShut {NoStop}%
\bibitem [{\citenamefont {Leeb}\ \emph {et~al.}(2024)\citenamefont {Leeb}, \citenamefont {Mook}, \citenamefont {{\v{S}}mejkal},\ and\ \citenamefont {Knolle}}]{leeb2024spontaneous}%
  \BibitemOpen
  \bibfield  {author} {\bibinfo {author} {\bibfnamefont {V.}~\bibnamefont {Leeb}}, \bibinfo {author} {\bibfnamefont {A.}~\bibnamefont {Mook}}, \bibinfo {author} {\bibfnamefont {L.}~\bibnamefont {{\v{S}}mejkal}}, \ and\ \bibinfo {author} {\bibfnamefont {J.}~\bibnamefont {Knolle}},\ }\href@noop {} {\bibfield  {journal} {\bibinfo  {journal} {Physical Review Letters}\ }\textbf {\bibinfo {volume} {132}},\ \bibinfo {pages} {236701} (\bibinfo {year} {2024})}\BibitemShut {NoStop}%
\bibitem [{\citenamefont {Milivojevi{\'c}}\ \emph {et~al.}(2024)\citenamefont {Milivojevi{\'c}}, \citenamefont {Orozovi{\'c}}, \citenamefont {Picozzi}, \citenamefont {Gmitra},\ and\ \citenamefont {Stavri{\'c}}}]{milivojevic2024interplay}%
  \BibitemOpen
  \bibfield  {author} {\bibinfo {author} {\bibfnamefont {M.}~\bibnamefont {Milivojevi{\'c}}}, \bibinfo {author} {\bibfnamefont {M.}~\bibnamefont {Orozovi{\'c}}}, \bibinfo {author} {\bibfnamefont {S.}~\bibnamefont {Picozzi}}, \bibinfo {author} {\bibfnamefont {M.}~\bibnamefont {Gmitra}}, \ and\ \bibinfo {author} {\bibfnamefont {S.}~\bibnamefont {Stavri{\'c}}},\ }\href@noop {} {\bibfield  {journal} {\bibinfo  {journal} {2D Materials}\ }\textbf {\bibinfo {volume} {11}},\ \bibinfo {pages} {035025} (\bibinfo {year} {2024})}\BibitemShut {NoStop}%
\bibitem [{\citenamefont {Yuan}\ \emph {et~al.}(2021)\citenamefont {Yuan}, \citenamefont {Wang}, \citenamefont {Luo},\ and\ \citenamefont {Zunger}}]{yuan2021prediction}%
  \BibitemOpen
  \bibfield  {author} {\bibinfo {author} {\bibfnamefont {L.-D.}\ \bibnamefont {Yuan}}, \bibinfo {author} {\bibfnamefont {Z.}~\bibnamefont {Wang}}, \bibinfo {author} {\bibfnamefont {J.-W.}\ \bibnamefont {Luo}}, \ and\ \bibinfo {author} {\bibfnamefont {A.}~\bibnamefont {Zunger}},\ }\href@noop {} {\bibfield  {journal} {\bibinfo  {journal} {Physical Review Materials}\ }\textbf {\bibinfo {volume} {5}},\ \bibinfo {pages} {014409} (\bibinfo {year} {2021})}\BibitemShut {NoStop}%
\bibitem [{\citenamefont {Yuan}\ \emph {et~al.}(2024)\citenamefont {Yuan}, \citenamefont {Georgescu},\ and\ \citenamefont {Rondinelli}}]{yuan2024non}%
  \BibitemOpen
  \bibfield  {author} {\bibinfo {author} {\bibfnamefont {L.-D.}\ \bibnamefont {Yuan}}, \bibinfo {author} {\bibfnamefont {A.~B.}\ \bibnamefont {Georgescu}}, \ and\ \bibinfo {author} {\bibfnamefont {J.~M.}\ \bibnamefont {Rondinelli}},\ }\href@noop {} {\bibfield  {journal} {\bibinfo  {journal} {arXiv preprint arXiv:2402.14321}\ } (\bibinfo {year} {2024})}\BibitemShut {NoStop}%
\bibitem [{\citenamefont {Nagaosa}\ \emph {et~al.}(2010)\citenamefont {Nagaosa}, \citenamefont {Sinova}, \citenamefont {Onoda}, \citenamefont {MacDonald},\ and\ \citenamefont {Ong}}]{nagaosa2010anomalous}%
  \BibitemOpen
  \bibfield  {author} {\bibinfo {author} {\bibfnamefont {N.}~\bibnamefont {Nagaosa}}, \bibinfo {author} {\bibfnamefont {J.}~\bibnamefont {Sinova}}, \bibinfo {author} {\bibfnamefont {S.}~\bibnamefont {Onoda}}, \bibinfo {author} {\bibfnamefont {A.~H.}\ \bibnamefont {MacDonald}}, \ and\ \bibinfo {author} {\bibfnamefont {N.~P.}\ \bibnamefont {Ong}},\ }\href@noop {} {\bibfield  {journal} {\bibinfo  {journal} {Reviews of modern physics}\ }\textbf {\bibinfo {volume} {82}},\ \bibinfo {pages} {1539} (\bibinfo {year} {2010})}\BibitemShut {NoStop}%
\bibitem [{\citenamefont {Chen}\ \emph {et~al.}(2014)\citenamefont {Chen}, \citenamefont {Niu},\ and\ \citenamefont {MacDonald}}]{chen2014anomalous}%
  \BibitemOpen
  \bibfield  {author} {\bibinfo {author} {\bibfnamefont {H.}~\bibnamefont {Chen}}, \bibinfo {author} {\bibfnamefont {Q.}~\bibnamefont {Niu}}, \ and\ \bibinfo {author} {\bibfnamefont {A.~H.}\ \bibnamefont {MacDonald}},\ }\href@noop {} {\bibfield  {journal} {\bibinfo  {journal} {Physical review letters}\ }\textbf {\bibinfo {volume} {112}},\ \bibinfo {pages} {017205} (\bibinfo {year} {2014})}\BibitemShut {NoStop}%
\bibitem [{\citenamefont {K{\"u}bler}\ and\ \citenamefont {Felser}(2014)}]{kubler2014non}%
  \BibitemOpen
  \bibfield  {author} {\bibinfo {author} {\bibfnamefont {J.}~\bibnamefont {K{\"u}bler}}\ and\ \bibinfo {author} {\bibfnamefont {C.}~\bibnamefont {Felser}},\ }\href@noop {} {\bibfield  {journal} {\bibinfo  {journal} {Europhysics Letters}\ }\textbf {\bibinfo {volume} {108}},\ \bibinfo {pages} {67001} (\bibinfo {year} {2014})}\BibitemShut {NoStop}%
\bibitem [{\citenamefont {Nakatsuji}\ \emph {et~al.}(2015)\citenamefont {Nakatsuji}, \citenamefont {Kiyohara},\ and\ \citenamefont {Higo}}]{nakatsuji2015large}%
  \BibitemOpen
  \bibfield  {author} {\bibinfo {author} {\bibfnamefont {S.}~\bibnamefont {Nakatsuji}}, \bibinfo {author} {\bibfnamefont {N.}~\bibnamefont {Kiyohara}}, \ and\ \bibinfo {author} {\bibfnamefont {T.}~\bibnamefont {Higo}},\ }\href@noop {} {\bibfield  {journal} {\bibinfo  {journal} {Nature}\ }\textbf {\bibinfo {volume} {527}},\ \bibinfo {pages} {212} (\bibinfo {year} {2015})}\BibitemShut {NoStop}%
\bibitem [{\citenamefont {S{\"u}rgers}\ \emph {et~al.}(2016)\citenamefont {S{\"u}rgers}, \citenamefont {Kittler}, \citenamefont {Wolf},\ and\ \citenamefont {L{\"o}hneysen}}]{surgers2016anomalous}%
  \BibitemOpen
  \bibfield  {author} {\bibinfo {author} {\bibfnamefont {C.}~\bibnamefont {S{\"u}rgers}}, \bibinfo {author} {\bibfnamefont {W.}~\bibnamefont {Kittler}}, \bibinfo {author} {\bibfnamefont {T.}~\bibnamefont {Wolf}}, \ and\ \bibinfo {author} {\bibfnamefont {H.~v.}\ \bibnamefont {L{\"o}hneysen}},\ }\href@noop {} {\bibfield  {journal} {\bibinfo  {journal} {AIP Advances}\ }\textbf {\bibinfo {volume} {6}} (\bibinfo {year} {2016})}\BibitemShut {NoStop}%
\bibitem [{\citenamefont {Boldrin}\ \emph {et~al.}(2019)\citenamefont {Boldrin}, \citenamefont {Samathrakis}, \citenamefont {Zemen}, \citenamefont {Mihai}, \citenamefont {Zou}, \citenamefont {Johnson}, \citenamefont {Esser}, \citenamefont {McComb}, \citenamefont {Petrov}, \citenamefont {Zhang} \emph {et~al.}}]{boldrin2019anomalous}%
  \BibitemOpen
  \bibfield  {author} {\bibinfo {author} {\bibfnamefont {D.}~\bibnamefont {Boldrin}}, \bibinfo {author} {\bibfnamefont {I.}~\bibnamefont {Samathrakis}}, \bibinfo {author} {\bibfnamefont {J.}~\bibnamefont {Zemen}}, \bibinfo {author} {\bibfnamefont {A.}~\bibnamefont {Mihai}}, \bibinfo {author} {\bibfnamefont {B.}~\bibnamefont {Zou}}, \bibinfo {author} {\bibfnamefont {F.}~\bibnamefont {Johnson}}, \bibinfo {author} {\bibfnamefont {B.~D.}\ \bibnamefont {Esser}}, \bibinfo {author} {\bibfnamefont {D.~W.}\ \bibnamefont {McComb}}, \bibinfo {author} {\bibfnamefont {P.~K.}\ \bibnamefont {Petrov}}, \bibinfo {author} {\bibfnamefont {H.}~\bibnamefont {Zhang}},  \emph {et~al.},\ }\href@noop {} {\bibfield  {journal} {\bibinfo  {journal} {Physical Review Materials}\ }\textbf {\bibinfo {volume} {3}},\ \bibinfo {pages} {094409} (\bibinfo {year} {2019})}\BibitemShut {NoStop}%
\bibitem [{\citenamefont {{\v{S}}mejkal}\ \emph {et~al.}(2022{\natexlab{c}})\citenamefont {{\v{S}}mejkal}, \citenamefont {MacDonald}, \citenamefont {Sinova}, \citenamefont {Nakatsuji},\ and\ \citenamefont {Jungwirth}}]{vsmejkal2022anomalous}%
  \BibitemOpen
  \bibfield  {author} {\bibinfo {author} {\bibfnamefont {L.}~\bibnamefont {{\v{S}}mejkal}}, \bibinfo {author} {\bibfnamefont {A.~H.}\ \bibnamefont {MacDonald}}, \bibinfo {author} {\bibfnamefont {J.}~\bibnamefont {Sinova}}, \bibinfo {author} {\bibfnamefont {S.}~\bibnamefont {Nakatsuji}}, \ and\ \bibinfo {author} {\bibfnamefont {T.}~\bibnamefont {Jungwirth}},\ }\href@noop {} {\bibfield  {journal} {\bibinfo  {journal} {Nature Reviews Materials}\ }\textbf {\bibinfo {volume} {7}},\ \bibinfo {pages} {482} (\bibinfo {year} {2022}{\natexlab{c}})}\BibitemShut {NoStop}%
\bibitem [{\citenamefont {Attias}\ \emph {et~al.}(2024)\citenamefont {Attias}, \citenamefont {Levchenko},\ and\ \citenamefont {Khodas}}]{attias2024intrinsic}%
  \BibitemOpen
  \bibfield  {author} {\bibinfo {author} {\bibfnamefont {L.}~\bibnamefont {Attias}}, \bibinfo {author} {\bibfnamefont {A.}~\bibnamefont {Levchenko}}, \ and\ \bibinfo {author} {\bibfnamefont {M.}~\bibnamefont {Khodas}},\ }\href@noop {} {\bibfield  {journal} {\bibinfo  {journal} {Physical Review B}\ }\textbf {\bibinfo {volume} {110}},\ \bibinfo {pages} {094425} (\bibinfo {year} {2024})}\BibitemShut {NoStop}%
\bibitem [{\citenamefont {Leivisk{\"a}}\ \emph {et~al.}(2024)\citenamefont {Leivisk{\"a}}, \citenamefont {Rial}, \citenamefont {Bad'ura}, \citenamefont {Seeger}, \citenamefont {Kounta}, \citenamefont {Beckert}, \citenamefont {Kriegner}, \citenamefont {Joumard}, \citenamefont {Schmoranzerov{\'a}}, \citenamefont {Sinova} \emph {et~al.}}]{leiviska2024anisotropy}%
  \BibitemOpen
  \bibfield  {author} {\bibinfo {author} {\bibfnamefont {M.}~\bibnamefont {Leivisk{\"a}}}, \bibinfo {author} {\bibfnamefont {J.}~\bibnamefont {Rial}}, \bibinfo {author} {\bibfnamefont {A.}~\bibnamefont {Bad'ura}}, \bibinfo {author} {\bibfnamefont {R.~L.}\ \bibnamefont {Seeger}}, \bibinfo {author} {\bibfnamefont {I.}~\bibnamefont {Kounta}}, \bibinfo {author} {\bibfnamefont {S.}~\bibnamefont {Beckert}}, \bibinfo {author} {\bibfnamefont {D.}~\bibnamefont {Kriegner}}, \bibinfo {author} {\bibfnamefont {I.}~\bibnamefont {Joumard}}, \bibinfo {author} {\bibfnamefont {E.}~\bibnamefont {Schmoranzerov{\'a}}}, \bibinfo {author} {\bibfnamefont {J.}~\bibnamefont {Sinova}},  \emph {et~al.},\ }\href@noop {} {\bibfield  {journal} {\bibinfo  {journal} {Physical Review B}\ }\textbf {\bibinfo {volume} {109}},\ \bibinfo {pages} {224430} (\bibinfo {year} {2024})}\BibitemShut {NoStop}%
\bibitem [{\citenamefont {Feng}\ \emph {et~al.}(2022)\citenamefont {Feng}, \citenamefont {Zhou}, \citenamefont {{\v{S}}mejkal}, \citenamefont {Wu}, \citenamefont {Zhu}, \citenamefont {Guo}, \citenamefont {Gonz{\'a}lez-Hern{\'a}ndez}, \citenamefont {Wang}, \citenamefont {Yan}, \citenamefont {Qin} \emph {et~al.}}]{feng2022anomalous}%
  \BibitemOpen
  \bibfield  {author} {\bibinfo {author} {\bibfnamefont {Z.}~\bibnamefont {Feng}}, \bibinfo {author} {\bibfnamefont {X.}~\bibnamefont {Zhou}}, \bibinfo {author} {\bibfnamefont {L.}~\bibnamefont {{\v{S}}mejkal}}, \bibinfo {author} {\bibfnamefont {L.}~\bibnamefont {Wu}}, \bibinfo {author} {\bibfnamefont {Z.}~\bibnamefont {Zhu}}, \bibinfo {author} {\bibfnamefont {H.}~\bibnamefont {Guo}}, \bibinfo {author} {\bibfnamefont {R.}~\bibnamefont {Gonz{\'a}lez-Hern{\'a}ndez}}, \bibinfo {author} {\bibfnamefont {X.}~\bibnamefont {Wang}}, \bibinfo {author} {\bibfnamefont {H.}~\bibnamefont {Yan}}, \bibinfo {author} {\bibfnamefont {P.}~\bibnamefont {Qin}},  \emph {et~al.},\ }\href@noop {} {\bibfield  {journal} {\bibinfo  {journal} {Nature Electronics}\ }\textbf {\bibinfo {volume} {5}},\ \bibinfo {pages} {735} (\bibinfo {year} {2022})}\BibitemShut {NoStop}%
\bibitem [{\citenamefont {Tschirner}\ \emph {et~al.}(2023)\citenamefont {Tschirner}, \citenamefont {Ke{\ss}ler}, \citenamefont {Gonzalez~Betancourt}, \citenamefont {Kotte}, \citenamefont {Kriegner}, \citenamefont {B{\"u}chner}, \citenamefont {Dufouleur}, \citenamefont {Kamp}, \citenamefont {Jovic}, \citenamefont {Smejkal} \emph {et~al.}}]{tschirner2023saturation}%
  \BibitemOpen
  \bibfield  {author} {\bibinfo {author} {\bibfnamefont {T.}~\bibnamefont {Tschirner}}, \bibinfo {author} {\bibfnamefont {P.}~\bibnamefont {Ke{\ss}ler}}, \bibinfo {author} {\bibfnamefont {R.~D.}\ \bibnamefont {Gonzalez~Betancourt}}, \bibinfo {author} {\bibfnamefont {T.}~\bibnamefont {Kotte}}, \bibinfo {author} {\bibfnamefont {D.}~\bibnamefont {Kriegner}}, \bibinfo {author} {\bibfnamefont {B.}~\bibnamefont {B{\"u}chner}}, \bibinfo {author} {\bibfnamefont {J.}~\bibnamefont {Dufouleur}}, \bibinfo {author} {\bibfnamefont {M.}~\bibnamefont {Kamp}}, \bibinfo {author} {\bibfnamefont {V.}~\bibnamefont {Jovic}}, \bibinfo {author} {\bibfnamefont {L.}~\bibnamefont {Smejkal}},  \emph {et~al.},\ }\href@noop {} {\bibfield  {journal} {\bibinfo  {journal} {APL Materials}\ }\textbf {\bibinfo {volume} {11}} (\bibinfo {year} {2023})}\BibitemShut {NoStop}%
\bibitem [{\citenamefont {Jin}\ \emph {et~al.}(2024)\citenamefont {Jin}, \citenamefont {Tan}, \citenamefont {Gong},\ and\ \citenamefont {Wang}}]{jin2024anomalous}%
  \BibitemOpen
  \bibfield  {author} {\bibinfo {author} {\bibfnamefont {H.}~\bibnamefont {Jin}}, \bibinfo {author} {\bibfnamefont {Z.}~\bibnamefont {Tan}}, \bibinfo {author} {\bibfnamefont {Z.}~\bibnamefont {Gong}}, \ and\ \bibinfo {author} {\bibfnamefont {J.}~\bibnamefont {Wang}},\ }\href@noop {} {\bibfield  {journal} {\bibinfo  {journal} {Physical Review B}\ }\textbf {\bibinfo {volume} {110}},\ \bibinfo {pages} {155125} (\bibinfo {year} {2024})}\BibitemShut {NoStop}%
\bibitem [{\citenamefont {Betancourt}\ \emph {et~al.}(2023)\citenamefont {Betancourt}, \citenamefont {Zub{\'a}{\v{c}}}, \citenamefont {Gonzalez-Hernandez}, \citenamefont {Geishendorf}, \citenamefont {{\v{S}}ob{\'a}{\v{n}}}, \citenamefont {Springholz}, \citenamefont {Olejn{\'\i}k}, \citenamefont {{\v{S}}mejkal}, \citenamefont {Sinova}, \citenamefont {Jungwirth} \emph {et~al.}}]{betancourt2023spontaneous}%
  \BibitemOpen
  \bibfield  {author} {\bibinfo {author} {\bibfnamefont {R.~G.}\ \bibnamefont {Betancourt}}, \bibinfo {author} {\bibfnamefont {J.}~\bibnamefont {Zub{\'a}{\v{c}}}}, \bibinfo {author} {\bibfnamefont {R.}~\bibnamefont {Gonzalez-Hernandez}}, \bibinfo {author} {\bibfnamefont {K.}~\bibnamefont {Geishendorf}}, \bibinfo {author} {\bibfnamefont {Z.}~\bibnamefont {{\v{S}}ob{\'a}{\v{n}}}}, \bibinfo {author} {\bibfnamefont {G.}~\bibnamefont {Springholz}}, \bibinfo {author} {\bibfnamefont {K.}~\bibnamefont {Olejn{\'\i}k}}, \bibinfo {author} {\bibfnamefont {L.}~\bibnamefont {{\v{S}}mejkal}}, \bibinfo {author} {\bibfnamefont {J.}~\bibnamefont {Sinova}}, \bibinfo {author} {\bibfnamefont {T.}~\bibnamefont {Jungwirth}},  \emph {et~al.},\ }\href@noop {} {\bibfield  {journal} {\bibinfo  {journal} {Physical Review Letters}\ }\textbf {\bibinfo {volume} {130}},\ \bibinfo {pages} {036702} (\bibinfo {year} {2023})}\BibitemShut {NoStop}%
\bibitem [{\citenamefont {Reichlov{\'a}}\ \emph {et~al.}(2024)\citenamefont {Reichlov{\'a}}, \citenamefont {Seeger}, \citenamefont {Gonz{\'a}lez-Hern{\'a}ndez}, \citenamefont {Kounta}, \citenamefont {Schlitz}, \citenamefont {Kriegner}, \citenamefont {Ritzinger}, \citenamefont {Lammel}, \citenamefont {Leivisk{\"a}}, \citenamefont {Pet{\v{r}}{\'\i}{\v{c}}ek} \emph {et~al.}}]{reichlova2024macroscopic}%
  \BibitemOpen
  \bibfield  {author} {\bibinfo {author} {\bibfnamefont {H.}~\bibnamefont {Reichlov{\'a}}}, \bibinfo {author} {\bibfnamefont {R.~L.}\ \bibnamefont {Seeger}}, \bibinfo {author} {\bibfnamefont {R.}~\bibnamefont {Gonz{\'a}lez-Hern{\'a}ndez}}, \bibinfo {author} {\bibfnamefont {I.}~\bibnamefont {Kounta}}, \bibinfo {author} {\bibfnamefont {R.}~\bibnamefont {Schlitz}}, \bibinfo {author} {\bibfnamefont {D.}~\bibnamefont {Kriegner}}, \bibinfo {author} {\bibfnamefont {P.}~\bibnamefont {Ritzinger}}, \bibinfo {author} {\bibfnamefont {M.}~\bibnamefont {Lammel}}, \bibinfo {author} {\bibfnamefont {M.}~\bibnamefont {Leivisk{\"a}}}, \bibinfo {author} {\bibfnamefont {V.}~\bibnamefont {Pet{\v{r}}{\'\i}{\v{c}}ek}},  \emph {et~al.},\ }\href@noop {} {\  (\bibinfo {year} {2024})}\BibitemShut {NoStop}%
\bibitem [{\citenamefont {Kluczyk}\ \emph {et~al.}(2024)\citenamefont {Kluczyk}, \citenamefont {Gas}, \citenamefont {Grzybowski}, \citenamefont {Skupi{\'n}ski}, \citenamefont {Borysiewicz}, \citenamefont {F{\k{a}}s}, \citenamefont {Suffczy{\'n}ski}, \citenamefont {Domagala}, \citenamefont {Grasza}, \citenamefont {Mycielski} \emph {et~al.}}]{kluczyk2024coexistence}%
  \BibitemOpen
\bibfield  {author} {\bibinfo {author} {\bibfnamefont {K.}~\bibnamefont {Kluczyk}}, \bibinfo {author} {\bibfnamefont {K.}~\bibnamefont {Gas}}, \bibinfo {author} {\bibfnamefont {M.}~\bibnamefont {Grzybowski}}, \bibinfo {author} {\bibfnamefont {P.}~\bibnamefont {Skupi{\'n}ski}}, \bibinfo {author} {\bibfnamefont {M.}~\bibnamefont {Borysiewicz}}, \bibinfo {author} {\bibfnamefont {T.}~\bibnamefont {F{{a}}s}}, \bibinfo {author} {\bibfnamefont {J.}~\bibnamefont {Suffczy{\'n}ski}}, \bibinfo {author} {\bibfnamefont {J.}~\bibnamefont {Domagala}}, \bibinfo {author} {\bibfnamefont {K.}~\bibnamefont {Grasza}}, \bibinfo {author} {\bibfnamefont {A.}~\bibnamefont {Mycielski}},\ }\href@noop {} {\bibfield  {journal} {\bibinfo  {journal} {Physical Review B}\ }\textbf {\bibinfo {volume} {110}},\ \bibinfo {pages} {155201} (\bibinfo {year} {2024})}\BibitemShut {NoStop}%
\bibitem [{\citenamefont {Khalili~Amiri}\ \emph {et~al.}(2024)\citenamefont {Khalili~Amiri}, \citenamefont {Phatak},\ and\ \citenamefont {Finocchio}}]{khalili2024prospects}%
  \BibitemOpen
  \bibfield  {author} {\bibinfo {author} {\bibfnamefont {P.}~\bibnamefont {Khalili~Amiri}}, \bibinfo {author} {\bibfnamefont {C.}~\bibnamefont {Phatak}}, \ and\ \bibinfo {author} {\bibfnamefont {G.}~\bibnamefont {Finocchio}},\ }\href@noop {} {\bibfield  {journal} {\bibinfo  {journal} {Annual Review of Materials Research}\ }\textbf {\bibinfo {volume} {54}} (\bibinfo {year} {2024})}\BibitemShut {NoStop}%
\bibitem [{\citenamefont {Jungwirth}\ \emph {et~al.}(2016)\citenamefont {Jungwirth}, \citenamefont {Marti}, \citenamefont {Wadley},\ and\ \citenamefont {Wunderlich}}]{jungwirth2016antiferromagnetic}%
  \BibitemOpen
  \bibfield  {author} {\bibinfo {author} {\bibfnamefont {T.}~\bibnamefont {Jungwirth}}, \bibinfo {author} {\bibfnamefont {X.}~\bibnamefont {Marti}}, \bibinfo {author} {\bibfnamefont {P.}~\bibnamefont {Wadley}}, \ and\ \bibinfo {author} {\bibfnamefont {J.}~\bibnamefont {Wunderlich}},\ }\href@noop {} {\bibfield  {journal} {\bibinfo  {journal} {Nature nanotechnology}\ }\textbf {\bibinfo {volume} {11}},\ \bibinfo {pages} {231} (\bibinfo {year} {2016})}\BibitemShut {NoStop}%
\bibitem [{\citenamefont {{\v{S}}mejkal}\ \emph {et~al.}(2018)\citenamefont {{\v{S}}mejkal}, \citenamefont {Mokrousov}, \citenamefont {Yan},\ and\ \citenamefont {MacDonald}}]{vsmejkal2018topological}%
  \BibitemOpen
  \bibfield  {author} {\bibinfo {author} {\bibfnamefont {L.}~\bibnamefont {{\v{S}}mejkal}}, \bibinfo {author} {\bibfnamefont {Y.}~\bibnamefont {Mokrousov}}, \bibinfo {author} {\bibfnamefont {B.}~\bibnamefont {Yan}}, \ and\ \bibinfo {author} {\bibfnamefont {A.~H.}\ \bibnamefont {MacDonald}},\ }\href@noop {} {\bibfield  {journal} {\bibinfo  {journal} {Nature physics}\ }\textbf {\bibinfo {volume} {14}},\ \bibinfo {pages} {242} (\bibinfo {year} {2018})}\BibitemShut {NoStop}%
\bibitem [{\citenamefont {Baltz}\ \emph {et~al.}(2018)\citenamefont {Baltz}, \citenamefont {Manchon}, \citenamefont {Tsoi}, \citenamefont {Moriyama}, \citenamefont {Ono},\ and\ \citenamefont {Tserkovnyak}}]{baltz2018antiferromagnetic}%
  \BibitemOpen
  \bibfield  {author} {\bibinfo {author} {\bibfnamefont {V.}~\bibnamefont {Baltz}}, \bibinfo {author} {\bibfnamefont {A.}~\bibnamefont {Manchon}}, \bibinfo {author} {\bibfnamefont {M.}~\bibnamefont {Tsoi}}, \bibinfo {author} {\bibfnamefont {T.}~\bibnamefont {Moriyama}}, \bibinfo {author} {\bibfnamefont {T.}~\bibnamefont {Ono}}, \ and\ \bibinfo {author} {\bibfnamefont {Y.}~\bibnamefont {Tserkovnyak}},\ }\href@noop {} {\bibfield  {journal} {\bibinfo  {journal} {Reviews of Modern Physics}\ }\textbf {\bibinfo {volume} {90}},\ \bibinfo {pages} {015005} (\bibinfo {year} {2018})}\BibitemShut {NoStop}%
\bibitem [{\citenamefont {Zhou}\ \emph {et~al.}(2024)\citenamefont {Zhou}, \citenamefont {Feng}, \citenamefont {Zhang}, \citenamefont {{\v{S}}mejkal}, \citenamefont {Sinova}, \citenamefont {Mokrousov},\ and\ \citenamefont {Yao}}]{zhou2024crystal}%
  \BibitemOpen
  \bibfield  {author} {\bibinfo {author} {\bibfnamefont {X.}~\bibnamefont {Zhou}}, \bibinfo {author} {\bibfnamefont {W.}~\bibnamefont {Feng}}, \bibinfo {author} {\bibfnamefont {R.-W.}\ \bibnamefont {Zhang}}, \bibinfo {author} {\bibfnamefont {L.}~\bibnamefont {{\v{S}}mejkal}}, \bibinfo {author} {\bibfnamefont {J.}~\bibnamefont {Sinova}}, \bibinfo {author} {\bibfnamefont {Y.}~\bibnamefont {Mokrousov}}, \ and\ \bibinfo {author} {\bibfnamefont {Y.}~\bibnamefont {Yao}},\ }\href@noop {} {\bibfield  {journal} {\bibinfo  {journal} {Physical review letters}\ }\textbf {\bibinfo {volume} {132}},\ \bibinfo {pages} {056701} (\bibinfo {year} {2024})}\BibitemShut {NoStop}%
\bibitem [{\citenamefont {Hasan}\ \emph {et~al.}(2021)\citenamefont {Hasan}, \citenamefont {Chang}, \citenamefont {Belopolski}, \citenamefont {Bian}, \citenamefont {Xu},\ and\ \citenamefont {Yin}}]{hasan2021weyl}%
  \BibitemOpen
  \bibfield  {author} {\bibinfo {author} {\bibfnamefont {M.~Z.}\ \bibnamefont {Hasan}}, \bibinfo {author} {\bibfnamefont {G.}~\bibnamefont {Chang}}, \bibinfo {author} {\bibfnamefont {I.}~\bibnamefont {Belopolski}}, \bibinfo {author} {\bibfnamefont {G.}~\bibnamefont {Bian}}, \bibinfo {author} {\bibfnamefont {S.-Y.}\ \bibnamefont {Xu}}, \ and\ \bibinfo {author} {\bibfnamefont {J.-X.}\ \bibnamefont {Yin}},\ }\href@noop {} {\bibfield  {journal} {\bibinfo  {journal} {Nature Reviews Materials}\ }\textbf {\bibinfo {volume} {6}},\ \bibinfo {pages} {784} (\bibinfo {year} {2021})}\BibitemShut {NoStop}%
\bibitem [{\citenamefont {Bradlyn}\ \emph {et~al.}(2016)\citenamefont {Bradlyn}, \citenamefont {Cano}, \citenamefont {Wang}, \citenamefont {Vergniory}, \citenamefont {Felser}, \citenamefont {Cava},\ and\ \citenamefont {Bernevig}}]{bradlyn2016beyond}%
  \BibitemOpen
  \bibfield  {author} {\bibinfo {author} {\bibfnamefont {B.}~\bibnamefont {Bradlyn}}, \bibinfo {author} {\bibfnamefont {J.}~\bibnamefont {Cano}}, \bibinfo {author} {\bibfnamefont {Z.}~\bibnamefont {Wang}}, \bibinfo {author} {\bibfnamefont {M.}~\bibnamefont {Vergniory}}, \bibinfo {author} {\bibfnamefont {C.}~\bibnamefont {Felser}}, \bibinfo {author} {\bibfnamefont {R.~J.}\ \bibnamefont {Cava}}, \ and\ \bibinfo {author} {\bibfnamefont {B.~A.}\ \bibnamefont {Bernevig}},\ }\href@noop {} {\bibfield  {journal} {\bibinfo  {journal} {Science}\ }\textbf {\bibinfo {volume} {353}},\ \bibinfo {pages} {aaf5037} (\bibinfo {year} {2016})}\BibitemShut {NoStop}%
\bibitem [{\citenamefont {Ke{\ss}ler}\ \emph {et~al.}(2024)\citenamefont {Ke{\ss}ler}, \citenamefont {Garcia-Gassull}, \citenamefont {Suter}, \citenamefont {Prokscha}, \citenamefont {Salman}, \citenamefont {Khalyavin}, \citenamefont {Manuel}, \citenamefont {Orlandi}, \citenamefont {Mazin}, \citenamefont {Valent{\'\i}} \emph {et~al.}}]{kessler2024absence}%
  \BibitemOpen
  \bibfield  {author} {\bibinfo {author} {\bibfnamefont {P.}~\bibnamefont {Ke{\ss}ler}}, \bibinfo {author} {\bibfnamefont {L.}~\bibnamefont {Garcia-Gassull}}, \bibinfo {author} {\bibfnamefont {A.}~\bibnamefont {Suter}}, \bibinfo {author} {\bibfnamefont {T.}~\bibnamefont {Prokscha}}, \bibinfo {author} {\bibfnamefont {Z.}~\bibnamefont {Salman}}, \bibinfo {author} {\bibfnamefont {D.}~\bibnamefont {Khalyavin}}, \bibinfo {author} {\bibfnamefont {P.}~\bibnamefont {Manuel}}, \bibinfo {author} {\bibfnamefont {F.}~\bibnamefont {Orlandi}}, \bibinfo {author} {\bibfnamefont {I.~I.}\ \bibnamefont {Mazin}}, \bibinfo {author} {\bibfnamefont {R.}~\bibnamefont {Valent{\'\i}}},  \emph {et~al.},\ }\href@noop {} {\bibfield  {journal} {\bibinfo  {journal} {npj Spintronics}\ }\textbf {\bibinfo {volume} {2}},\ \bibinfo {pages} {50} (\bibinfo {year} {2024})}\BibitemShut {NoStop}%
\bibitem [{\citenamefont {Lovesey}\ \emph {et~al.}(2023)\citenamefont {Lovesey}, \citenamefont {Khalyavin},\ and\ \citenamefont {Van Der~Laan}}]{lovesey2023templates}%
  \BibitemOpen
  \bibfield  {author} {\bibinfo {author} {\bibfnamefont {S.}~\bibnamefont {Lovesey}}, \bibinfo {author} {\bibfnamefont {D.}~\bibnamefont {Khalyavin}}, \ and\ \bibinfo {author} {\bibfnamefont {G.}~\bibnamefont {Van Der~Laan}},\ }\href@noop {} {\bibfield  {journal} {\bibinfo  {journal} {Physical Review B}\ }\textbf {\bibinfo {volume} {108}},\ \bibinfo {pages} {174437} (\bibinfo {year} {2023})}\BibitemShut {NoStop}%
\bibitem [{\citenamefont {Hariki}\ \emph {et~al.}(2024{\natexlab{a}})\citenamefont {Hariki}, \citenamefont {Dal~Din}, \citenamefont {Amin}, \citenamefont {Yamaguchi}, \citenamefont {Badura}, \citenamefont {Kriegner}, \citenamefont {Edmonds}, \citenamefont {Campion}, \citenamefont {Wadley}, \citenamefont {Backes} \emph {et~al.}}]{hariki2024x}%
  \BibitemOpen
  \bibfield  {author} {\bibinfo {author} {\bibfnamefont {A.}~\bibnamefont {Hariki}}, \bibinfo {author} {\bibfnamefont {A.}~\bibnamefont {Dal~Din}}, \bibinfo {author} {\bibfnamefont {O.}~\bibnamefont {Amin}}, \bibinfo {author} {\bibfnamefont {T.}~\bibnamefont {Yamaguchi}}, \bibinfo {author} {\bibfnamefont {A.}~\bibnamefont {Badura}}, \bibinfo {author} {\bibfnamefont {D.}~\bibnamefont {Kriegner}}, \bibinfo {author} {\bibfnamefont {K.}~\bibnamefont {Edmonds}}, \bibinfo {author} {\bibfnamefont {R.}~\bibnamefont {Campion}}, \bibinfo {author} {\bibfnamefont {P.}~\bibnamefont {Wadley}}, \bibinfo {author} {\bibfnamefont {D.}~\bibnamefont {Backes}},  \emph {et~al.},\ }\href@noop {} {\bibfield  {journal} {\bibinfo  {journal} {Physical Review Letters}\ }\textbf {\bibinfo {volume} {132}},\ \bibinfo {pages} {176701} (\bibinfo {year} {2024}{\natexlab{a}})}\BibitemShut {NoStop}%
\bibitem [{\citenamefont {Hariki}\ \emph {et~al.}(2024{\natexlab{b}})\citenamefont {Hariki}, \citenamefont {Okauchi}, \citenamefont {Takahashi},\ and\ \citenamefont {Kune{\v{s}}}}]{hariki2024determination}%
  \BibitemOpen
  \bibfield  {author} {\bibinfo {author} {\bibfnamefont {A.}~\bibnamefont {Hariki}}, \bibinfo {author} {\bibfnamefont {T.}~\bibnamefont {Okauchi}}, \bibinfo {author} {\bibfnamefont {Y.}~\bibnamefont {Takahashi}}, \ and\ \bibinfo {author} {\bibfnamefont {J.}~\bibnamefont {Kune{\v{s}}}},\ }\href@noop {} {\bibfield  {journal} {\bibinfo  {journal} {Physical Review B}\ }\textbf {\bibinfo {volume} {110}},\ \bibinfo {pages} {L100402} (\bibinfo {year} {2024}{\natexlab{b}})}\BibitemShut {NoStop}%
\bibitem [{\citenamefont {Lee}\ \emph {et~al.}(2024)\citenamefont {Lee}, \citenamefont {Lee}, \citenamefont {Jung}, \citenamefont {Jung}, \citenamefont {Kim}, \citenamefont {Lee}, \citenamefont {Seok}, \citenamefont {Kim}, \citenamefont {Park}, \citenamefont {{\v{S}}mejkal} \emph {et~al.}}]{lee2024broken}%
  \BibitemOpen
  \bibfield  {author} {\bibinfo {author} {\bibfnamefont {S.}~\bibnamefont {Lee}}, \bibinfo {author} {\bibfnamefont {S.}~\bibnamefont {Lee}}, \bibinfo {author} {\bibfnamefont {S.}~\bibnamefont {Jung}}, \bibinfo {author} {\bibfnamefont {J.}~\bibnamefont {Jung}}, \bibinfo {author} {\bibfnamefont {D.}~\bibnamefont {Kim}}, \bibinfo {author} {\bibfnamefont {Y.}~\bibnamefont {Lee}}, \bibinfo {author} {\bibfnamefont {B.}~\bibnamefont {Seok}}, \bibinfo {author} {\bibfnamefont {J.}~\bibnamefont {Kim}}, \bibinfo {author} {\bibfnamefont {B.~G.}\ \bibnamefont {Park}}, \bibinfo {author} {\bibfnamefont {L.}~\bibnamefont {{\v{S}}mejkal}},  \emph {et~al.},\ }\href@noop {} {\bibfield  {journal} {\bibinfo  {journal} {Physical Review Letters}\ }\textbf {\bibinfo {volume} {132}},\ \bibinfo {pages} {036702} (\bibinfo {year} {2024})}\BibitemShut {NoStop}%
\bibitem [{\citenamefont {Liu}\ \emph {et~al.}(2024{\natexlab{a}})\citenamefont {Liu}, \citenamefont {Zhan}, \citenamefont {Li}, \citenamefont {Liu}, \citenamefont {Cheng}, \citenamefont {Shi}, \citenamefont {Deng}, \citenamefont {Zhang}, \citenamefont {Li}, \citenamefont {Ding} \emph {et~al.}}]{liu2024absence}%
  \BibitemOpen
  \bibfield  {author} {\bibinfo {author} {\bibfnamefont {J.}~\bibnamefont {Liu}}, \bibinfo {author} {\bibfnamefont {J.}~\bibnamefont {Zhan}}, \bibinfo {author} {\bibfnamefont {T.}~\bibnamefont {Li}}, \bibinfo {author} {\bibfnamefont {J.}~\bibnamefont {Liu}}, \bibinfo {author} {\bibfnamefont {S.}~\bibnamefont {Cheng}}, \bibinfo {author} {\bibfnamefont {Y.}~\bibnamefont {Shi}}, \bibinfo {author} {\bibfnamefont {L.}~\bibnamefont {Deng}}, \bibinfo {author} {\bibfnamefont {M.}~\bibnamefont {Zhang}}, \bibinfo {author} {\bibfnamefont {C.}~\bibnamefont {Li}}, \bibinfo {author} {\bibfnamefont {J.}~\bibnamefont {Ding}},  \emph {et~al.},\ }\href@noop {} {\bibfield  {journal} {\bibinfo  {journal} {Physical Review Letters}\ }\textbf {\bibinfo {volume} {133}},\ \bibinfo {pages} {176401} (\bibinfo {year} {2024}{\natexlab{a}})}\BibitemShut {NoStop}%
\bibitem [{\citenamefont {Krempask{\`y}}\ \emph {et~al.}(2024)\citenamefont {Krempask{\`y}}, \citenamefont {{\v{S}}mejkal}, \citenamefont {D’souza}, \citenamefont {Hajlaoui}, \citenamefont {Springholz}, \citenamefont {Uhl{\'\i}{\v{r}}ov{\'a}}, \citenamefont {Alarab}, \citenamefont {Constantinou}, \citenamefont {Strocov}, \citenamefont {Usanov} \emph {et~al.}}]{krempasky2024altermagnetic}%
  \BibitemOpen
  \bibfield  {author} {\bibinfo {author} {\bibfnamefont {J.}~\bibnamefont {Krempask{\`y}}}, \bibinfo {author} {\bibfnamefont {L.}~\bibnamefont {{\v{S}}mejkal}}, \bibinfo {author} {\bibfnamefont {S.}~\bibnamefont {D’souza}}, \bibinfo {author} {\bibfnamefont {M.}~\bibnamefont {Hajlaoui}}, \bibinfo {author} {\bibfnamefont {G.}~\bibnamefont {Springholz}}, \bibinfo {author} {\bibfnamefont {K.}~\bibnamefont {Uhl{\'\i}{\v{r}}ov{\'a}}}, \bibinfo {author} {\bibfnamefont {F.}~\bibnamefont {Alarab}}, \bibinfo {author} {\bibfnamefont {P.}~\bibnamefont {Constantinou}}, \bibinfo {author} {\bibfnamefont {V.}~\bibnamefont {Strocov}}, \bibinfo {author} {\bibfnamefont {D.}~\bibnamefont {Usanov}},  \emph {et~al.},\ }\href@noop {} {\bibfield  {journal} {\bibinfo  {journal} {Nature}\ }\textbf {\bibinfo {volume} {626}},\ \bibinfo {pages} {517} (\bibinfo {year} {2024})}\BibitemShut {NoStop}%
\bibitem [{\citenamefont {Osumi}\ \emph {et~al.}(2024)\citenamefont {Osumi}, \citenamefont {Souma}, \citenamefont {Aoyama}, \citenamefont {Yamauchi}, \citenamefont {Honma}, \citenamefont {Nakayama}, \citenamefont {Takahashi}, \citenamefont {Ohgushi},\ and\ \citenamefont {Sato}}]{osumi2024observation}%
  \BibitemOpen
  \bibfield  {author} {\bibinfo {author} {\bibfnamefont {T.}~\bibnamefont {Osumi}}, \bibinfo {author} {\bibfnamefont {S.}~\bibnamefont {Souma}}, \bibinfo {author} {\bibfnamefont {T.}~\bibnamefont {Aoyama}}, \bibinfo {author} {\bibfnamefont {K.}~\bibnamefont {Yamauchi}}, \bibinfo {author} {\bibfnamefont {A.}~\bibnamefont {Honma}}, \bibinfo {author} {\bibfnamefont {K.}~\bibnamefont {Nakayama}}, \bibinfo {author} {\bibfnamefont {T.}~\bibnamefont {Takahashi}}, \bibinfo {author} {\bibfnamefont {K.}~\bibnamefont {Ohgushi}}, \ and\ \bibinfo {author} {\bibfnamefont {T.}~\bibnamefont {Sato}},\ }\href@noop {} {\bibfield  {journal} {\bibinfo  {journal} {Physical Review B}\ }\textbf {\bibinfo {volume} {109}},\ \bibinfo {pages} {115102} (\bibinfo {year} {2024})}\BibitemShut {NoStop}%
\bibitem [{\citenamefont {Zhu}\ \emph {et~al.}(2024)\citenamefont {Zhu}, \citenamefont {Chen}, \citenamefont {Liu}, \citenamefont {Liu}, \citenamefont {Liu}, \citenamefont {Zha}, \citenamefont {Qu}, \citenamefont {Hong}, \citenamefont {Li}, \citenamefont {Jiang} \emph {et~al.}}]{zhu2024observation}%
  \BibitemOpen
  \bibfield  {author} {\bibinfo {author} {\bibfnamefont {Y.-P.}\ \bibnamefont {Zhu}}, \bibinfo {author} {\bibfnamefont {X.}~\bibnamefont {Chen}}, \bibinfo {author} {\bibfnamefont {X.-R.}\ \bibnamefont {Liu}}, \bibinfo {author} {\bibfnamefont {Y.}~\bibnamefont {Liu}}, \bibinfo {author} {\bibfnamefont {P.}~\bibnamefont {Liu}}, \bibinfo {author} {\bibfnamefont {H.}~\bibnamefont {Zha}}, \bibinfo {author} {\bibfnamefont {G.}~\bibnamefont {Qu}}, \bibinfo {author} {\bibfnamefont {C.}~\bibnamefont {Hong}}, \bibinfo {author} {\bibfnamefont {J.}~\bibnamefont {Li}}, \bibinfo {author} {\bibfnamefont {Z.}~\bibnamefont {Jiang}},  \emph {et~al.},\ }\href@noop {} {\bibfield  {journal} {\bibinfo  {journal} {Nature}\ }\textbf {\bibinfo {volume} {626}},\ \bibinfo {pages} {523} (\bibinfo {year} {2024})}\BibitemShut {NoStop}%
\bibitem [{\citenamefont {Reimers}\ \emph {et~al.}(2024)\citenamefont {Reimers}, \citenamefont {Odenbreit}, \citenamefont {{\v{S}}mejkal}, \citenamefont {Strocov}, \citenamefont {Constantinou}, \citenamefont {Hellenes}, \citenamefont {Jaeschke~Ubiergo}, \citenamefont {Campos}, \citenamefont {Bharadwaj}, \citenamefont {Chakraborty} \emph {et~al.}}]{reimers2024direct}%
  \BibitemOpen
  \bibfield  {author} {\bibinfo {author} {\bibfnamefont {S.}~\bibnamefont {Reimers}}, \bibinfo {author} {\bibfnamefont {L.}~\bibnamefont {Odenbreit}}, \bibinfo {author} {\bibfnamefont {L.}~\bibnamefont {{\v{S}}mejkal}}, \bibinfo {author} {\bibfnamefont {V.~N.}\ \bibnamefont {Strocov}}, \bibinfo {author} {\bibfnamefont {P.}~\bibnamefont {Constantinou}}, \bibinfo {author} {\bibfnamefont {A.~B.}\ \bibnamefont {Hellenes}}, \bibinfo {author} {\bibfnamefont {R.}~\bibnamefont {Jaeschke~Ubiergo}}, \bibinfo {author} {\bibfnamefont {W.~H.}\ \bibnamefont {Campos}}, \bibinfo {author} {\bibfnamefont {V.~K.}\ \bibnamefont {Bharadwaj}}, \bibinfo {author} {\bibfnamefont {A.}~\bibnamefont {Chakraborty}},  \emph {et~al.},\ }\href@noop {} {\bibfield  {journal} {\bibinfo  {journal} {Nature Communications}\ }\textbf {\bibinfo {volume} {15}},\ \bibinfo {pages} {2116} (\bibinfo {year} {2024})}\BibitemShut {NoStop}%
\bibitem [{\citenamefont {Ding}\ \emph {et~al.}(2024)\citenamefont {Ding}, \citenamefont {Jiang}, \citenamefont {Chen}, \citenamefont {Tao}, \citenamefont {Liu}, \citenamefont {Li}, \citenamefont {Liu}, \citenamefont {Sun}, \citenamefont {Cheng}, \citenamefont {Liu} \emph {et~al.}}]{ding2024large}%
  \BibitemOpen
  \bibfield  {author} {\bibinfo {author} {\bibfnamefont {J.}~\bibnamefont {Ding}}, \bibinfo {author} {\bibfnamefont {Z.}~\bibnamefont {Jiang}}, \bibinfo {author} {\bibfnamefont {X.}~\bibnamefont {Chen}}, \bibinfo {author} {\bibfnamefont {Z.}~\bibnamefont {Tao}}, \bibinfo {author} {\bibfnamefont {Z.}~\bibnamefont {Liu}}, \bibinfo {author} {\bibfnamefont {T.}~\bibnamefont {Li}}, \bibinfo {author} {\bibfnamefont {J.}~\bibnamefont {Liu}}, \bibinfo {author} {\bibfnamefont {J.}~\bibnamefont {Sun}}, \bibinfo {author} {\bibfnamefont {J.}~\bibnamefont {Cheng}}, \bibinfo {author} {\bibfnamefont {J.}~\bibnamefont {Liu}},  \emph {et~al.},\ }\href@noop {} {\bibfield  {journal} {\bibinfo  {journal} {Physical Review Letters}\ }\textbf {\bibinfo {volume} {133}},\ \bibinfo {pages} {206401} (\bibinfo {year} {2024})}\BibitemShut {NoStop}%
\bibitem [{\citenamefont {Lin}\ \emph {et~al.}(2024)\citenamefont {Lin}, \citenamefont {Chen}, \citenamefont {Lu}, \citenamefont {Liang}, \citenamefont {Feng}, \citenamefont {Yamagami}, \citenamefont {Osiecki}, \citenamefont {Leandersson}, \citenamefont {Thiagarajan}, \citenamefont {Liu} \emph {et~al.}}]{lin2024observation}%
  \BibitemOpen
  \bibfield  {author} {\bibinfo {author} {\bibfnamefont {Z.}~\bibnamefont {Lin}}, \bibinfo {author} {\bibfnamefont {D.}~\bibnamefont {Chen}}, \bibinfo {author} {\bibfnamefont {W.}~\bibnamefont {Lu}}, \bibinfo {author} {\bibfnamefont {X.}~\bibnamefont {Liang}}, \bibinfo {author} {\bibfnamefont {S.}~\bibnamefont {Feng}}, \bibinfo {author} {\bibfnamefont {K.}~\bibnamefont {Yamagami}}, \bibinfo {author} {\bibfnamefont {J.}~\bibnamefont {Osiecki}}, \bibinfo {author} {\bibfnamefont {M.}~\bibnamefont {Leandersson}}, \bibinfo {author} {\bibfnamefont {B.}~\bibnamefont {Thiagarajan}}, \bibinfo {author} {\bibfnamefont {J.}~\bibnamefont {Liu}},  \emph {et~al.},\ }\href@noop {} {\bibfield  {journal} {\bibinfo  {journal} {arXiv preprint arXiv:2402.04995}\ } (\bibinfo {year} {2024})}\BibitemShut {NoStop}%
\bibitem [{\citenamefont {Ahn}\ \emph {et~al.}(2019)\citenamefont {Ahn}, \citenamefont {Hariki}, \citenamefont {Lee},\ and\ \citenamefont {Kune{\v{s}}}}]{ahn2019antiferromagnetism}%
  \BibitemOpen
  \bibfield  {author} {\bibinfo {author} {\bibfnamefont {K.-H.}\ \bibnamefont {Ahn}}, \bibinfo {author} {\bibfnamefont {A.}~\bibnamefont {Hariki}}, \bibinfo {author} {\bibfnamefont {K.-W.}\ \bibnamefont {Lee}}, \ and\ \bibinfo {author} {\bibfnamefont {J.}~\bibnamefont {Kune{\v{s}}}},\ }\href@noop {} {\bibfield  {journal} {\bibinfo  {journal} {Physical Review B}\ }\textbf {\bibinfo {volume} {99}},\ \bibinfo {pages} {184432} (\bibinfo {year} {2019})}\BibitemShut {NoStop}%
\bibitem [{\citenamefont {Bhowal}\ and\ \citenamefont {Spaldin}(2024)}]{bhowal2022magnetic}%
  \BibitemOpen
  \bibfield  {author} {\bibinfo {author} {\bibfnamefont {S.}~\bibnamefont {Bhowal}}\ and\ \bibinfo {author} {\bibfnamefont {N.~A.}\ \bibnamefont {Spaldin}},\ }\href {\doibase 10.1103/PhysRevX.14.011019} {\bibfield  {journal} {\bibinfo  {journal} {Phys. Rev. X}\ }\textbf {\bibinfo {volume} {14}},\ \bibinfo {pages} {011019} (\bibinfo {year} {2024})}\BibitemShut {NoStop}%
\bibitem [{\citenamefont {Zeng}\ \emph {et~al.}(2024)\citenamefont {Zeng}, \citenamefont {Zhu}, \citenamefont {Zhu}, \citenamefont {Liu}, \citenamefont {Ma}, \citenamefont {Hao}, \citenamefont {Liu}, \citenamefont {Qu}, \citenamefont {Yang}, \citenamefont {Jiang} \emph {et~al.}}]{zeng2024observation}%
  \BibitemOpen
  \bibfield  {author} {\bibinfo {author} {\bibfnamefont {M.}~\bibnamefont {Zeng}}, \bibinfo {author} {\bibfnamefont {M.-Y.}\ \bibnamefont {Zhu}}, \bibinfo {author} {\bibfnamefont {Y.-P.}\ \bibnamefont {Zhu}}, \bibinfo {author} {\bibfnamefont {X.-R.}\ \bibnamefont {Liu}}, \bibinfo {author} {\bibfnamefont {X.-M.}\ \bibnamefont {Ma}}, \bibinfo {author} {\bibfnamefont {Y.-J.}\ \bibnamefont {Hao}}, \bibinfo {author} {\bibfnamefont {P.}~\bibnamefont {Liu}}, \bibinfo {author} {\bibfnamefont {G.}~\bibnamefont {Qu}}, \bibinfo {author} {\bibfnamefont {Y.}~\bibnamefont {Yang}}, \bibinfo {author} {\bibfnamefont {Z.}~\bibnamefont {Jiang}},  \emph {et~al.},\ }\href@noop {} {\bibfield  {journal} {\bibinfo  {journal} {Advanced Science}\ ,\ \bibinfo {pages} {2406529}} (\bibinfo {year} {2024})}\BibitemShut {NoStop}%
\bibitem [{\citenamefont {Liu}\ \emph {et~al.}(2024{\natexlab{b}})\citenamefont {Liu}, \citenamefont {Kang}, \citenamefont {Wang}, \citenamefont {Gao}, \citenamefont {Qi}, \citenamefont {Zhao},\ and\ \citenamefont {Jiang}}]{liu2024inverse}%
  \BibitemOpen
  \bibfield  {author} {\bibinfo {author} {\bibfnamefont {Q.}~\bibnamefont {Liu}}, \bibinfo {author} {\bibfnamefont {J.}~\bibnamefont {Kang}}, \bibinfo {author} {\bibfnamefont {P.}~\bibnamefont {Wang}}, \bibinfo {author} {\bibfnamefont {W.}~\bibnamefont {Gao}}, \bibinfo {author} {\bibfnamefont {Y.}~\bibnamefont {Qi}}, \bibinfo {author} {\bibfnamefont {J.}~\bibnamefont {Zhao}}, \ and\ \bibinfo {author} {\bibfnamefont {X.}~\bibnamefont {Jiang}},\ }\href@noop {} {\bibfield  {journal} {\bibinfo  {journal} {Advanced Functional Materials}\ ,\ \bibinfo {pages} {2402080}} (\bibinfo {year} {2024}{\natexlab{b}})}\BibitemShut {NoStop}%
\bibitem [{\citenamefont {Ryden}\ and\ \citenamefont {Lawson}(1970)}]{ryden1970magnetic}%
  \BibitemOpen
  \bibfield  {author} {\bibinfo {author} {\bibfnamefont {W.}~\bibnamefont {Ryden}}\ and\ \bibinfo {author} {\bibfnamefont {A.}~\bibnamefont {Lawson}},\ }\href@noop {} {\bibfield  {journal} {\bibinfo  {journal} {The Journal of Chemical Physics}\ }\textbf {\bibinfo {volume} {52}},\ \bibinfo {pages} {6058} (\bibinfo {year} {1970})}\BibitemShut {NoStop}%
\bibitem [{\citenamefont {Berlijn}\ \emph {et~al.}(2017)\citenamefont {Berlijn}, \citenamefont {Snijders}, \citenamefont {Delaire}, \citenamefont {Zhou}, \citenamefont {Maier}, \citenamefont {Cao}, \citenamefont {Chi}, \citenamefont {Matsuda}, \citenamefont {Wang}, \citenamefont {Koehler} \emph {et~al.}}]{berlijn2017itinerant}%
  \BibitemOpen
  \bibfield  {author} {\bibinfo {author} {\bibfnamefont {T.}~\bibnamefont {Berlijn}}, \bibinfo {author} {\bibfnamefont {P.~C.}\ \bibnamefont {Snijders}}, \bibinfo {author} {\bibfnamefont {O.}~\bibnamefont {Delaire}}, \bibinfo {author} {\bibfnamefont {H.-D.}\ \bibnamefont {Zhou}}, \bibinfo {author} {\bibfnamefont {T.~A.}\ \bibnamefont {Maier}}, \bibinfo {author} {\bibfnamefont {H.-B.}\ \bibnamefont {Cao}}, \bibinfo {author} {\bibfnamefont {S.-X.}\ \bibnamefont {Chi}}, \bibinfo {author} {\bibfnamefont {M.}~\bibnamefont {Matsuda}}, \bibinfo {author} {\bibfnamefont {Y.}~\bibnamefont {Wang}}, \bibinfo {author} {\bibfnamefont {M.~R.}\ \bibnamefont {Koehler}},  \emph {et~al.},\ }\href@noop {} {\bibfield  {journal} {\bibinfo  {journal} {Physical review letters}\ }\textbf {\bibinfo {volume} {118}},\ \bibinfo {pages} {077201} (\bibinfo {year} {2017})}\BibitemShut {NoStop}%
\bibitem [{\citenamefont {Zhu}\ \emph {et~al.}(2019)\citenamefont {Zhu}, \citenamefont {Strempfer}, \citenamefont {Rao}, \citenamefont {Occhialini}, \citenamefont {Pelliciari}, \citenamefont {Choi}, \citenamefont {Kawaguchi}, \citenamefont {You}, \citenamefont {Mitchell}, \citenamefont {Shao-Horn} \emph {et~al.}}]{zhu2019anomalous}%
  \BibitemOpen
  \bibfield  {author} {\bibinfo {author} {\bibfnamefont {Z.}~\bibnamefont {Zhu}}, \bibinfo {author} {\bibfnamefont {J.}~\bibnamefont {Strempfer}}, \bibinfo {author} {\bibfnamefont {R.}~\bibnamefont {Rao}}, \bibinfo {author} {\bibfnamefont {C.}~\bibnamefont {Occhialini}}, \bibinfo {author} {\bibfnamefont {J.}~\bibnamefont {Pelliciari}}, \bibinfo {author} {\bibfnamefont {Y.}~\bibnamefont {Choi}}, \bibinfo {author} {\bibfnamefont {T.}~\bibnamefont {Kawaguchi}}, \bibinfo {author} {\bibfnamefont {H.}~\bibnamefont {You}}, \bibinfo {author} {\bibfnamefont {J.}~\bibnamefont {Mitchell}}, \bibinfo {author} {\bibfnamefont {Y.}~\bibnamefont {Shao-Horn}},  \emph {et~al.},\ }\href@noop {} {\bibfield  {journal} {\bibinfo  {journal} {Physical review letters}\ }\textbf {\bibinfo {volume} {122}},\ \bibinfo {pages} {017202} (\bibinfo {year} {2019})}\BibitemShut {NoStop}%
\bibitem [{\citenamefont {Smolyanyuk}\ \emph {et~al.}(2024{\natexlab{a}})\citenamefont {Smolyanyuk}, \citenamefont {Mazin}, \citenamefont {Garcia-Gassull},\ and\ \citenamefont {Valent{\'\i}}}]{smolyanyuk2024fragility}%
  \BibitemOpen
  \bibfield  {author} {\bibinfo {author} {\bibfnamefont {A.}~\bibnamefont {Smolyanyuk}}, \bibinfo {author} {\bibfnamefont {I.~I.}\ \bibnamefont {Mazin}}, \bibinfo {author} {\bibfnamefont {L.}~\bibnamefont {Garcia-Gassull}}, \ and\ \bibinfo {author} {\bibfnamefont {R.}~\bibnamefont {Valent{\'\i}}},\ }\href@noop {} {\bibfield  {journal} {\bibinfo  {journal} {Physical Review B}\ }\textbf {\bibinfo {volume} {109}},\ \bibinfo {pages} {134424} (\bibinfo {year} {2024}{\natexlab{a}})}\BibitemShut {NoStop}%
\bibitem [{\citenamefont {Hiraishi}\ \emph {et~al.}(2024)\citenamefont {Hiraishi}, \citenamefont {Okabe}, \citenamefont {Koda}, \citenamefont {Kadono}, \citenamefont {Muroi}, \citenamefont {Hirai},\ and\ \citenamefont {Hiroi}}]{hiraishi2024nonmagnetic}%
  \BibitemOpen
  \bibfield  {author} {\bibinfo {author} {\bibfnamefont {M.}~\bibnamefont {Hiraishi}}, \bibinfo {author} {\bibfnamefont {H.}~\bibnamefont {Okabe}}, \bibinfo {author} {\bibfnamefont {A.}~\bibnamefont {Koda}}, \bibinfo {author} {\bibfnamefont {R.}~\bibnamefont {Kadono}}, \bibinfo {author} {\bibfnamefont {T.}~\bibnamefont {Muroi}}, \bibinfo {author} {\bibfnamefont {D.}~\bibnamefont {Hirai}}, \ and\ \bibinfo {author} {\bibfnamefont {Z.}~\bibnamefont {Hiroi}},\ }\href@noop {} {\bibfield  {journal} {\bibinfo  {journal} {Physical Review Letters}\ }\textbf {\bibinfo {volume} {132}},\ \bibinfo {pages} {166702} (\bibinfo {year} {2024})}\BibitemShut {NoStop}%
\bibitem [{\citenamefont {{\v{S}}mejkal}\ \emph {et~al.}(2023)\citenamefont {{\v{S}}mejkal}, \citenamefont {Marmodoro}, \citenamefont {Ahn}, \citenamefont {Gonz{\'a}lez-Hern{\'a}ndez}, \citenamefont {Turek}, \citenamefont {Mankovsky}, \citenamefont {Ebert}, \citenamefont {D’Souza}, \citenamefont {{\v{S}}ipr}, \citenamefont {Sinova} \emph {et~al.}}]{vsmejkal2023chiral}%
  \BibitemOpen
  \bibfield  {author} {\bibinfo {author} {\bibfnamefont {L.}~\bibnamefont {{\v{S}}mejkal}}, \bibinfo {author} {\bibfnamefont {A.}~\bibnamefont {Marmodoro}}, \bibinfo {author} {\bibfnamefont {K.-H.}\ \bibnamefont {Ahn}}, \bibinfo {author} {\bibfnamefont {R.}~\bibnamefont {Gonz{\'a}lez-Hern{\'a}ndez}}, \bibinfo {author} {\bibfnamefont {I.}~\bibnamefont {Turek}}, \bibinfo {author} {\bibfnamefont {S.}~\bibnamefont {Mankovsky}}, \bibinfo {author} {\bibfnamefont {H.}~\bibnamefont {Ebert}}, \bibinfo {author} {\bibfnamefont {S.~W.}\ \bibnamefont {D’Souza}}, \bibinfo {author} {\bibfnamefont {O.}~\bibnamefont {{\v{S}}ipr}}, \bibinfo {author} {\bibfnamefont {J.}~\bibnamefont {Sinova}},  \emph {et~al.},\ }\href@noop {} {\bibfield  {journal} {\bibinfo  {journal} {Physical Review Letters}\ }\textbf {\bibinfo {volume} {131}},\ \bibinfo {pages} {256703} (\bibinfo {year} {2023})}\BibitemShut {NoStop}%
\bibitem [{\citenamefont {Gohlke}\ \emph {et~al.}(2023)\citenamefont {Gohlke}, \citenamefont {Corticelli}, \citenamefont {Moessner}, \citenamefont {McClarty},\ and\ \citenamefont {Mook}}]{gohlke2023spurious}%
  \BibitemOpen
  \bibfield  {author} {\bibinfo {author} {\bibfnamefont {M.}~\bibnamefont {Gohlke}}, \bibinfo {author} {\bibfnamefont {A.}~\bibnamefont {Corticelli}}, \bibinfo {author} {\bibfnamefont {R.}~\bibnamefont {Moessner}}, \bibinfo {author} {\bibfnamefont {P.~A.}\ \bibnamefont {McClarty}}, \ and\ \bibinfo {author} {\bibfnamefont {A.}~\bibnamefont {Mook}},\ }\href@noop {} {\bibfield  {journal} {\bibinfo  {journal} {Physical Review Letters}\ }\textbf {\bibinfo {volume} {131}},\ \bibinfo {pages} {186702} (\bibinfo {year} {2023})}\BibitemShut {NoStop}%
\bibitem [{\citenamefont {Litvin}(1977)}]{litvin1977spin}%
  \BibitemOpen
  \bibfield  {author} {\bibinfo {author} {\bibfnamefont {D.~B.}\ \bibnamefont {Litvin}},\ }\href@noop {} {\bibfield  {journal} {\bibinfo  {journal} {Acta Crystallographica Section A: Crystal Physics, Diffraction, Theoretical and General Crystallography}\ }\textbf {\bibinfo {volume} {33}},\ \bibinfo {pages} {279} (\bibinfo {year} {1977})}\BibitemShut {NoStop}%
\bibitem [{\citenamefont {Litvin}\ and\ \citenamefont {Opechowski}(1974)}]{litvin1974spin}%
  \BibitemOpen
  \bibfield  {author} {\bibinfo {author} {\bibfnamefont {D.~B.}\ \bibnamefont {Litvin}}\ and\ \bibinfo {author} {\bibfnamefont {W.}~\bibnamefont {Opechowski}},\ }\href@noop {} {\bibfield  {journal} {\bibinfo  {journal} {Physica}\ }\textbf {\bibinfo {volume} {76}},\ \bibinfo {pages} {538} (\bibinfo {year} {1974})}\BibitemShut {NoStop}%
\bibitem [{\citenamefont {Litvin}(1980)}]{litvin1980wreath}%
  \BibitemOpen
  \bibfield  {author} {\bibinfo {author} {\bibfnamefont {D.~B.}\ \bibnamefont {Litvin}},\ }\href@noop {} {\bibfield  {journal} {\bibinfo  {journal} {Physical Review B}\ }\textbf {\bibinfo {volume} {21}},\ \bibinfo {pages} {3184} (\bibinfo {year} {1980})}\BibitemShut {NoStop}%
\bibitem [{\citenamefont {Smolyanyuk}\ \emph {et~al.}(2024{\natexlab{b}})\citenamefont {Smolyanyuk}, \citenamefont {{\v{S}}mejkal},\ and\ \citenamefont {Mazin}}]{smolyanyuk2024tool}%
  \BibitemOpen
  \bibfield  {author} {\bibinfo {author} {\bibfnamefont {A.}~\bibnamefont {Smolyanyuk}}, \bibinfo {author} {\bibfnamefont {L.}~\bibnamefont {{\v{S}}mejkal}}, \ and\ \bibinfo {author} {\bibfnamefont {I.~I.}\ \bibnamefont {Mazin}},\ }\href@noop {} {\bibfield  {journal} {\bibinfo  {journal} {SciPost Physics Codebases}\ ,\ \bibinfo {pages} {030}} (\bibinfo {year} {2024}{\natexlab{b}})}\BibitemShut {NoStop}%
\bibitem [{\citenamefont {Sukhachov}\ \emph {et~al.}(2024)\citenamefont {Sukhachov}, \citenamefont {Hodt},\ and\ \citenamefont {Linder}}]{sukhachov2024thermoelectric}%
  \BibitemOpen
  \bibfield  {author} {\bibinfo {author} {\bibfnamefont {P.~O.}\ \bibnamefont {Sukhachov}}, \bibinfo {author} {\bibfnamefont {E.~W.}\ \bibnamefont {Hodt}}, \ and\ \bibinfo {author} {\bibfnamefont {J.}~\bibnamefont {Linder}},\ }\href@noop {} {\bibfield  {journal} {\bibinfo  {journal} {Physical Review B}\ }\textbf {\bibinfo {volume} {110}},\ \bibinfo {pages} {094508} (\bibinfo {year} {2024})}\BibitemShut {NoStop}%
\bibitem [{\citenamefont {Han}\ \emph {et~al.}(2024)\citenamefont {Han}, \citenamefont {Fu}, \citenamefont {He}, \citenamefont {Zhu}, \citenamefont {Dai}, \citenamefont {Yang}, \citenamefont {Zhu}, \citenamefont {Bai}, \citenamefont {Chen}, \citenamefont {Wan} \emph {et~al.}}]{han2024observation}%
  \BibitemOpen
  \bibfield  {author} {\bibinfo {author} {\bibfnamefont {L.}~\bibnamefont {Han}}, \bibinfo {author} {\bibfnamefont {X.}~\bibnamefont {Fu}}, \bibinfo {author} {\bibfnamefont {W.}~\bibnamefont {He}}, \bibinfo {author} {\bibfnamefont {Y.}~\bibnamefont {Zhu}}, \bibinfo {author} {\bibfnamefont {J.}~\bibnamefont {Dai}}, \bibinfo {author} {\bibfnamefont {W.}~\bibnamefont {Yang}}, \bibinfo {author} {\bibfnamefont {W.}~\bibnamefont {Zhu}}, \bibinfo {author} {\bibfnamefont {H.}~\bibnamefont {Bai}}, \bibinfo {author} {\bibfnamefont {C.}~\bibnamefont {Chen}}, \bibinfo {author} {\bibfnamefont {C.}~\bibnamefont {Wan}},  \emph {et~al.},\ }\href@noop {} {\bibfield  {journal} {\bibinfo  {journal} {arXiv preprint arXiv:2403.13427}\ } (\bibinfo {year} {2024})}\BibitemShut {NoStop}%
\end{thebibliography}%
		
	\end{document}